%% file: Dohyun_IEEE_reference.tex
\documentclass[12pt, journal, onecolumn, draftclsnofoot]{IEEEtran} 

\input{Dohyun_input.tex}
\title{Joint Band Assignment and Beam Management using Hierarchical Reinforcement Learning for Multi-Band Communication}
\author{
Dohyun Kim,~\IEEEmembership{Graduate Student Member, IEEE},
Miguel R. Castellanos,~\IEEEmembership{Member, IEEE}, and Robert W. Heath Jr.,~\IEEEmembership{Fellow, IEEE}
\thanks{Dohyun Kim is with the Wireless Networking and Communications Group, the University of Texas at Austin, TX 78712-1687, USA (e-mail: dohyun.kim@utexas.edu). Miguel R. Castellanos and Robert W. Heath Jr. are with the Department of Electrical and Computer Engineering, North Carolina State University, 890 Oval Dr., Raleigh, NC 27606 USA (email: \{mrcastel, rwheathjr\}@ncsu.edu). This material is based upon work supported by the National Science Foundation under grant nos. NSF-ECCS-2153698, NSF-CCF-2225555, NSF-CNS-2147955  and is supported in part by funds from federal agency and industry partners as specified in the Resilient \& Intelligent NextG Systems (RINGS) program.}
}
\begin{document}
\maketitle
\begin{abstract}
Multi-band operation in wireless networks can improve data rates by leveraging the benefits of propagation in different frequency ranges. Distinctive beam management procedures in different bands complicate band assignment because they require considering not only the channel quality but also the associated beam management overhead. Reinforcement learning (RL) is a promising approach for multi-band operation as it enables the system to learn and adjust its behavior through environmental feedback. In this paper, we formulate a sequential decision problem to jointly perform band assignment and beam management. We propose a method based on hierarchical RL (HRL) to handle the complexity of the problem by separating the policies for band selection and beam management. We evaluate the proposed HRL-based algorithm on a realistic channel generated based on ray-tracing simulators. Our results show that the proposed approach outperforms traditional RL approaches in terms of reduced beam training overhead and increased data rates under a realistic vehicular channel.
\end{abstract}
%well-known for its recent success in addressing the overhead of beam training in millimeter-wave (mmWave) vehicular networks encountering the challenges in the mmWave band, such as mobility and blockage.
%The proposed algorithm accounts for differences in beam management and feedback in different bands.

%In the ongoing discussion on 5G-Advanced, multi-band operation in vehicular networks is considered an effective approach to address these challenges.

%%%%%%%%%%%%%%%%%%%%%%%%%%%%%%%%%%%%%%%%%%%%%%%%%%%%%%%%%%%%%
\section{Introduction}
%%%%%%%%%%%%%%%%%%%%%%%%%%%%%%%%%%%%%%%%%%%%%%%%%%%%%%%%%%%%%

Multi-band systems can achieve high data rates while maintaining coverage and reliability \cite{ZhaNiYou:Multi-band-cooperation-5G-HetNet:VTM19}. The integration of mmWave and sub-6 GHz transceivers means that a single device can leverage the high bandwidths and data rates of mmWave communication and the resilient and wide coverage of sub-6 GHz communication \cite{YanDinHua:ML-handover-sub6-mmWave-integrated-vehicular-network:TCOMM19}. 3GPP continues to work on multi-band in recent Release 18, aiming to extend technology such as the sidelink from frequency range 1 (FR1, 0.4 GHz – 7.1 GHz) to frequency range 2 (FR2, 24 GHz – 52 GHz) \cite{Lin:3GPP-Release-18-overview:CSM22}. Multi-band operations can be realized into two directions: simultaneous transmission and band assignment. Allowing simultaneous usage of bands in a single time slot offers greater data rate potential but higher radio-frequency (RF) complexity. In this paper, we focus on band assignment that refers to the selection of operating bands over time slots in a sequential manner. Band assignment can be understood as a subproblem of the frequency resource allocation in multi-band systems \cite{LopKlaHea:DRL-scheduling-multiband-massive-MIMO:Access22}.

Beam management establishes and maintains beamformed links and is a critical component of both mmWave and sub-6 GHz communication. In the mmWave band, beam management is used to overcome beam misalignment and outages caused by mobility and blockages \cite{MarAguGom:Tech-Model-challenge-mmWave-vehicular-comm:COMMM18}. The beam management procedure in sub-6 GHz 5G uses uses a precoder matrix indicator (PMI) codebook and feedback, categorized as Type-1 and Type-2. Type-1 codebooks, with shorter training overhead and predefined precoders per antenna geometry, are commonly used for spatial multiplexing compared to Type-2 codebooks. The overhead from beam management, which deteriorates the data rate, can be excessive when exhaustive beam alignment methods that use narrow beam codebooks \cite{GioPolZor:Beam_Management_3GPP_NR_mmWave_Tutorial:COMM18}. Reducing the overhead of the beam management procedure in mmWave wireless networks can be complicated when the dynamics of the node, environment and the network is highly variable, as in the case in 5G vehicular networks \cite{FazMalNor:Belief-propagation-mmWave-beam-coordination:22TWC}.

The overhead of beam management has been overlooked in existing solutions for band assignment \cite{LiJayBka:Band-selection-for-spectrum-sensing:TCOMM14,AreJayMac:MARL-cognitive-anti-jamming-band-selection:WCNC17,NajMacBec:Deep-learning-RF-VLC-band-selection-D2D:COMM'L20,BurWanAlg:Supervised-ML-band-assignment:22TCOMM}. Prior work has studied various objectives such as throughput maximization \cite{LiJayBka:Band-selection-for-spectrum-sensing:TCOMM14,BurWanAlg:Supervised-ML-band-assignment:22TCOMM}, jammer interference minimization \cite{AreJayMac:MARL-cognitive-anti-jamming-band-selection:WCNC17}, and outage ratio minimization \cite{NajMacBec:Deep-learning-RF-VLC-band-selection-D2D:COMM'L20}, but they typically assume instant evaluation of a selected band. When accounting for beam management, the objective of minimizing the beam management overhead becomes an additional factor that can influence band selection. In the case of throughput maximization, while the mmWave band offers high throughput, the sub-6 GHz band can be more favorable due to its shorter beam management overhead, despite having lower throughput. High mobility and non-line-of-sight (NLOS) operation are representative examples in which the sub-6 GHz band could outperform the mmWave band. This highlights the need for a joint formulation of band assignment and beam management, where the tradeoff between throughput and overhead is carefully balanced.

Recent work has shown that reinforcement learning (RL) is an effective framework to address the overhead of beam training in mmWave vehicular networks \cite{KimCasHea:Joint-relay-selection-beam-management:TVT23,SimKloHol:Online_Context_Aware_ML_mmWAVE_VANET:TNET18,HusMic:Dual-timescale-mmWave-beam-tracking-and-training:JSAC22}. The RL framework, including partially observable environments and contextual bandits in the wide sense, is capable of reducing control overhead by using the accumulated deployment history to effectively balance the exploration of new control actions with the exploitation of actions that have yielded the highest expected return in the past. In our prior work \cite{KimCasHea:Joint-relay-selection-beam-management:TVT23}, we have studied deep RL (DRL) with threshold-based actions to reduce beam training overhead in mmWave MIMO vehicular networks with relay selection. In \cite{SimKloHol:Online_Context_Aware_ML_mmWAVE_VANET:TNET18}, the incoming vehicle direction was used as input to apply contextual bandits for beam selection in mmWave vehicular networks. In \cite{HusMic:Dual-timescale-mmWave-beam-tracking-and-training:JSAC22}, an autoencoder was employed to predict vehicle mobility then to find beam training policies based on RL with partial observability. RL has also been used in band assignment, where WiFi traffic demands are learned from WiFi channel activity observations \cite{TanZhaLia:DRL-MAC-LTE-WiFi:TCOMM20}. Still, beam management in multi-band wireless networks can be challenging based on traditional RL approaches because beam training can only be performed in one band at a time and the sample efficiency in each band will be low. Furthermore, stationarity of the learning model may not hold in the joint band assignment and beam management problem due to the distinctive beam training procedures in the two bands.

%We propose hierarchical reinforcement learning (HRL), which is a recent extension in the RL framework, to improve sample efficiency by separately learning policies for tasks on different decision hierarchy levels. 
Hierarchical reinforcement learning (HRL), which is a recent extension in the RL framework, is a promising approach for addressing the joint band assignment and beam management problem. HRL improves sample efficiency compared to DRL by separately learning policies for tasks at different decision hierarchy levels. The idea of hierarchy resembles intelligence observed in nature, such as humans performing few-shot learning on complex tasks by goal-oriented compositional abstraction \cite{EppGumWer:Bio-inspired-HRL:NMI22}. One of the main benefits of exploiting hierarchy is that the shortened episodes, owing to the abstracted tasks, makes both exploration and learning easier \cite{NacGuLev:Data-efficient-HRL:NIPS18}. While HRL is still a relatively new approach in wireless communication, it has shown promising results outperforming the traditional DRL methods in resource allocation \cite{GenLiuWan:HRL-relay-selection-power-optimization:TCOMM21}, channel sensing \cite{LiuWuHe:Channel-sensing-HRL:Access21}, and scheduling \cite{RenNiuGui:UAV-scheduling-HRL:IOT22}. Relevant work on HRL applications in wireless communication, however, rely on discrete action spaces that can limit their generalization to real-world problems. For example, the discrete action space only represent the quantized transmission power constraint in the power allocation problem \cite{GenLiuWan:HRL-relay-selection-power-optimization:TCOMM21}. We employ HRL using continuous actions, which can improve the scalability of the learning algorithm and enhance its applicability to real-world deployments. %Given its potential to handle complex tasks with reduced learning samples, HRL is a promising approach for addressing the joint band assignment and beam management problem.

In this paper, we propose an HRL-based algorithm for joint band assignment and beam management that leverages the band characteristics of sub-6 GHz and mmWave. We presume the communication nodes employ codebook-based beamforming, co-located sub-6 GHz and mmWave arrays, and Orthogonal Frequency Division Multiplexing (OFDM). We also assume a fully digital sub-6 GHz array and a hybrid mmWave array with analog and digital beamformers. The system can either perform digital beam training at the sub-6 GHz band, analog beam training at the mmWave band, or digital beam training at the mmWave band. We assume perfect spectral efficiency feedback from the user to the base station without quantization or overhead in the digital beam training at the sub-6 GHz band and the analog beam training at the mmWave band. The feedback may be available through a dedicated channel in the unoccupied band or may be sent on the reverse link with reduced coding and spreading. For the digital beam training at the mmWave band, we assume quantized baseband effective channel feedback using a random vector quantization (RVQ) codebook. The algorithm employs two policies: an upper-level policy for band selection and a lower-level policy to determine the beam training method. The choice of beam training is guided by comparing the spectral efficiency feedback and two adaptive thresholds determined by the lower-level policy. We use one threshold to separate analog and digital beam training and the other threshold to decide between digital beam training and data transmission. The band selection is made by the upper-policy, which aggregates state, goal, and reward over an adaptive period. The HRL-based method uses the best known band until the spectral efficiency feedback deteriorates below the learned threshold, in which case the algorithm tries out different band or beam training indicated by the upper-level and lower-level policies.

We summarize our contributions as follows:
\begin{enumerate}
	\item We formulate a joint band assignment and beam management
problem for wireless networks operating on sub-6 GHz and mmWave that accounts for the effect of the beam management overhead on the cumulative data rate. We devise a hierarchical sequential decision-making
model of the joint band assignment and beam management
problem, avoiding the non-stationary Markov decision process (MDP) by separately learning policies for band selection and beam management.
	\item We propose an HRL-based algorithm to solve the
	joint band assignment and beam management problem.
	The proposed algorithm uses the spectral efficiency
	feedback from the receiver to learn thresholds that determines the beam training method.  
	\item We numerically evaluate the proposed algorithm compared to baseline learning algorithms on a realistic vehicular channel. The HRL-based proposed algorithm outperforms the heuristic owing to the reduced horizon for policy computation from abstracted subtasks.
\end{enumerate}

The rest of the paper is structured as follows. In \secref{sec:system_model}, we present the system model used to represent the multi-band wireless network. In \secref{sec:problem_definition}, we formulate the joint band assignment and beam management problem and discuss the challenges of designing an learning algorithm. In \secref{sec:action_masking_baseline}, we describe a DRL algorithm that can partially address the challenges of the joint band assignment and beam management problem. In \secref{sec:main_algorithm}, we develop an HRL algorithm to solve the joint band assignment and mode selection problem. In \secref{sec:experiments}, we numerically evaluate the proposed algorithm compared to baselines. Finally, we conclude the paper in \secref{sec:conclusion}.

We use the following notation throughout this paper: $\bA$ is a matrix, $\ba$ is a vector, $a$ is a scalar, and $\cA$ is a set. We denote $\ba^{\mathrm{T}}$ the transpose of $\ba$, $\ba^{*}$ the conjugate transpose, $\|\ba\|_{2}$ the 2-norm, and  $\|\ba\|_{\text{F}}$ the Frobenius norm. We underline the sub-6 GHz variables as $\underline{\ba}$ to distinguish them from mmWave.

%%%%%%%%%%%%%%%%%%%%%%%%%%%%%%%%%%%%%%%%%%%%%%%%%%%%%%%%%%%%%
\section{System model}\label{sec:system_model}
%%%%%%%%%%%%%%%%%%%%%%%%%%%%%%%%%%%%%%%%%%%%%%%%%%%%%%%%%%%%%

\begin{figure*}[h!]
	\centering
	\subfloat[]{\includegraphics[width=0.6\columnwidth,draft=false]{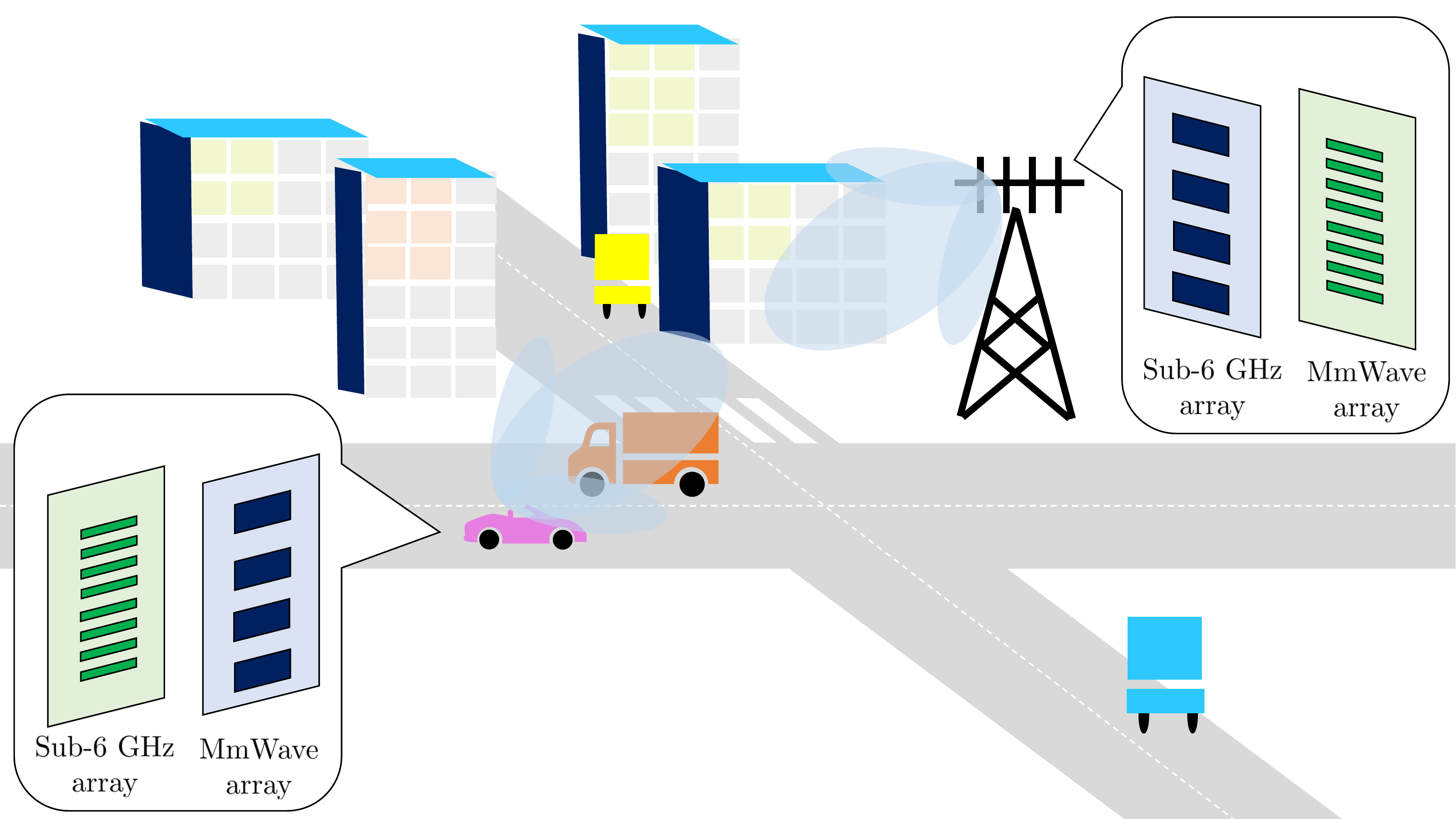}%
		\label{fig:ScenarioFigure1}
	}
	\hfil
	\subfloat[]{\includegraphics[width=0.6\columnwidth,draft=false]{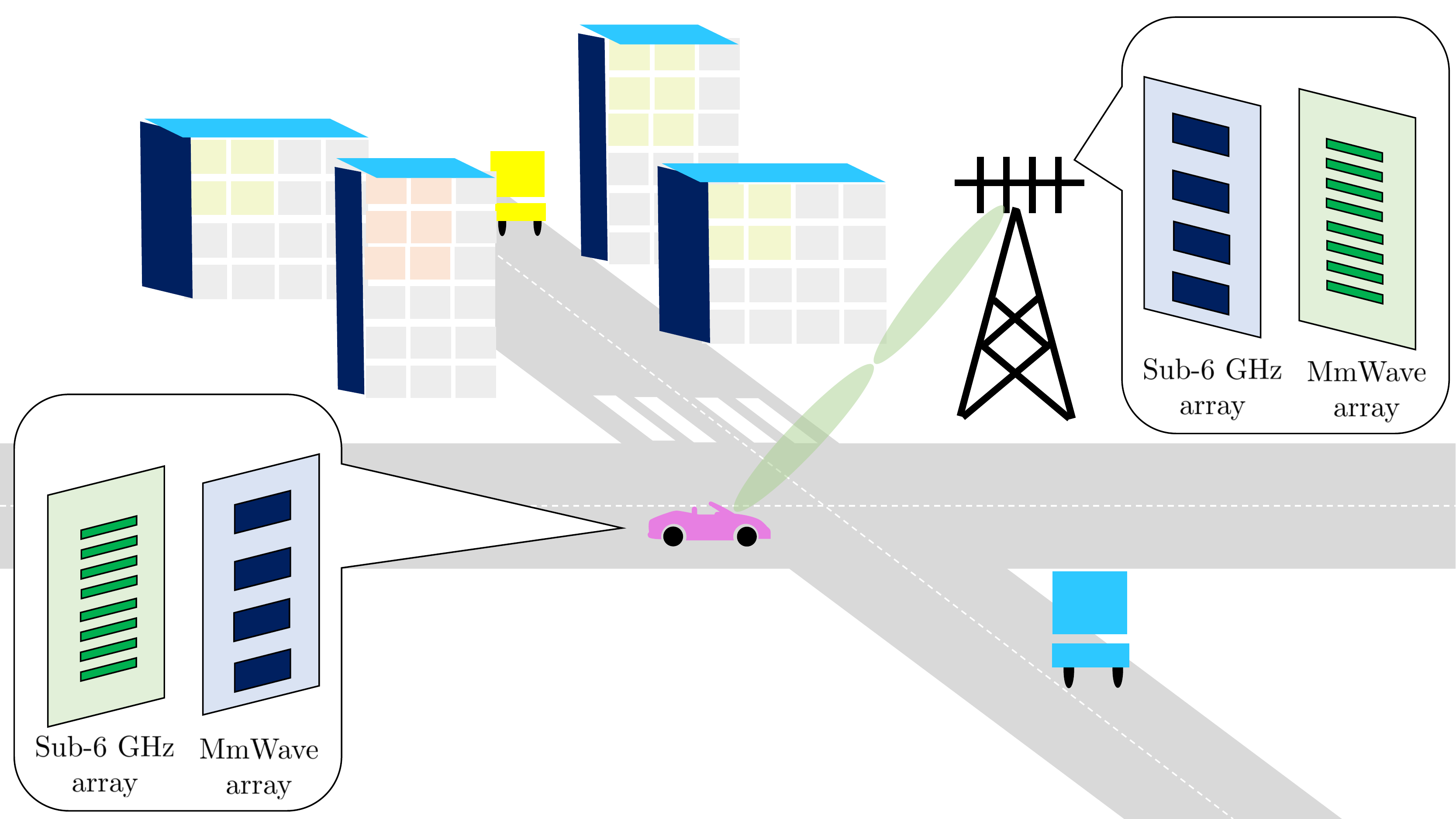}%
		\label{fig:ScenarioFigure2}
	}
	\caption{Illustration of an example system model showing two snapshots: (a) the base station {operates} on the sub-6 GHz band to serve the user due to a large truck posing as a mobile blockage, and (b) the base station {operates} on the mmWave band when LOS is available.}
	\label{fig:ScenarioFigure}
\end{figure*}

In this section, we describe the system model for a wireless network operating both on the sub-6 GHz and mmWave bands. As shown in \figref{fig:ScenarioFigure}, the system can operate on one band at a time. We assume that the communication nodes are equipped with co-located sub-6 GHz and mmWave arrays. We provide the signal model in the mmWave band in \secref{sec:mmWave_signal_model}. We then outline the codebooks and beam training procedure in the mmWave band in \secref{sec:beam_management}.  
We summarize the sub-6 GHz signal model and beam training process in  \secref{sec:sub6_system_model} and \secref{sec:sub6_beam_management}.

Consider a downlink scenario in a multi-band MIMO-OFDM wireless network, where a single base station serves a single mobile user. For each OFDM time frame, we assume the base station selects a transmission mode of either beam training or data transmission. We {also} assume the base station sends pilots only during beam training for $\BTlen$ discrete time slots. Whenever the mode is data transmission, the base station sends only data symbols for $\DTlen$ discrete time slots. The sequence of modes can be consecutive beam training, consecutive data transmissions, or alternating with an arbitrary number of consecutive modes. The band selection occurs when a new transmission mode is deployed. When the system uses the sub-6 GHz band, denoted by a binary variable $\bandassign=0$, the system operates over a bandwidth $\cmWavebandwidth$ with $\cmWavesubcarriernum$ subcarriers. Similarly, we use $\bandassign=1$ to denote that the system operates in the mmWave band. In this case, the system uses a bandwidth $\mmWavebandwidth$ with $\mmWavesubcarriernum$ subcarriers.

%===========================================================%
\subsection{Millimeter wave signal model}\label{sec:mmWave_signal_model}
%===========================================================%

In the mmWave band, we assume the system employs a fully connected hybrid beamforming architecture. We denote $\Nbs$ as the number of antennas and $\NBSRF$ as the number of RF chains at the base station. At the user, we denote $\userNr$ as the number of antennas and $\userNRF$ as the number of RF chains. The base station and the user communicate via $\userNS$ data streams, where $\NS\leq\NBSRF\leq\Nbs$ and $\userNS\leq\userNRF\leq\userNr$. For simplicity, we focus on a fully connected hybrid beamforming architecture in mmWave. Partially connected architectures can reduce power consumption and hardware cost, though beam training can be more complex as the whole channel may need to be reconstructed from subarray measurements. We leave the extension to a partially connected architecture for future work. 

At each OFDM time frame $\horizonindex$ {and subcarrier $\subcarrierindex$}, the base station sends a symbol vector $\usersymbol[\subcarrierindex,\horizonindex]$ of size $\NS \times 1$ to the user. The symbol vector is assumed to be normalized such that $\bbE[|\mathsf{\usersymbol}[\subcarrierindex,\horizonindex]|^{2}] = 1$. The base station precodes the symbol vector with {the} $\NBSRF \times \userNS$ frequency-selective baseband precoder $\userFBB[\subcarrierindex, \horizonindex]$ followed by the $\Nbs \times \NBSRF$ frequency-flat RF precoder $\FRF[\horizonindex]$. We assume the precoded signal propagates through a time-varying wideband channel model $\userchannel[\subcarrierindex, \horizonindex]$ with large-scale fading denoted as $\userlargescalefading$ and the noise denoted as $\usernoise[\subcarrierindex,\horizonindex]$. We assume the noise is independently and identically distributed (IID) following the distribution $\cN_{C}(0,\noiseSTD^{2})$. At the user, the received signal is processed with {the} $\userNr \times \userNRF$ frequency-flat RF {combiner} $\userWRF[\horizonindex]$ followed by the $\userNRF \times \userNS$ frequency-selective baseband {combiner} $\userWBB[\subcarrierindex, \horizonindex]$. We {set} power constraint on the base station by denoting $\userTransmitpower$ as the transmit power and {constraining} $\userFBB[\subcarrierindex, \horizonindex]$ such that $\|\FRF[\subcarrierindex,\horizonindex]\userFBB[\subcarrierindex, \horizonindex]\|_{F}^{2}=\userNS$. 

The end-to-end input-to-output relation in the mmWave band is
\begin{align}
\bm{\mathsf{y}}[\subcarrierindex, \horizonindex] &=\sqrt{\userTransmitpower\userlargescalefading} \userWBB^{*}[\subcarrierindex, \horizonindex] \userWRF^{*}[\horizonindex] \userchannel[\subcarrierindex,\horizonindex] \nonumber \\&\times \FRF[\horizonindex] \userFBB[\subcarrierindex, \horizonindex] \usersymbol[\subcarrierindex,\horizonindex]+\userWBB^{*}[\subcarrierindex, \horizonindex] \userWRF^{*}[\horizonindex] \usernoise[\subcarrierindex,\horizonindex].
\end{align}
We define the spectral efficiency {per the subcarrier $\subcarrierindex$} in the mmWave band as
\begin{align}
\userSE[\subcarrierindex,\horizonindex]&= \log\det \left(\mathbf{I}_{\userNS}+\userTransmitpower\userlargescalefading\noiseSTD^{-2} \userWBB^{*}[\subcarrierindex, \horizonindex]\userWRF^{*}[\horizonindex] \userchannel[\subcarrierindex, \horizonindex]
\right.\nonumber\\
&\left.\times
  \FRF[\horizonindex] \userFBB[\subcarrierindex, \horizonindex]\userFBB^{*}[\horizonindex] \FRF^{*}[\horizonindex] \userchannel^{*}[\subcarrierindex, \horizonindex] \userWRF[\horizonindex] \userWBB[\subcarrierindex, \horizonindex] \right)
\end{align}
Note that the signal-to-noise-ratio (SNR) prior to beamforming is $\userTransmitpower\userlargescalefading\noiseSTD^{-2}$ due to the normalization  $\bbE[|\mathsf{\usersymbol}[\subcarrierindex,\horizonindex]|^{2}] = 1$.

%===========================================================%
\subsection{Millimeter wave beam management procedure}\label{sec:beam_management}
%===========================================================%

In this section, we outline the beam management procedure {used in} the mmWave band. The purpose of the beam management procedure is to determine {the beamforming matrices---$\FRF[\horizonindex]$, $\userFBB[\subcarrierindex,\horizonindex]$, $\userWRF[\horizonindex]$, and $\userFBB[\subcarrierindex,\horizonindex]$---}adaptive to the dynamic channel conditions using feedback from the user to the base station. We assume that beam training can be split into two stages: analog and digital beam training. {The analog beam training is based on beam codebooks, such that the base station and the user select beam pairs. We further assume the analog beam training} involves exchanging synchronization signals (SSs) between the base station and the user \cite{GioPolZor:Beam_Management_3GPP_NR_mmWave_Tutorial:COMM18}. In the digital beam training, the user estimates the digital effective channel and computes the digital combiner then feeds back the quantized effective channel to the base station.

We first describe the analog beam training procedure in the mmWave band. 
Let us denote the base station analog codebook with size $\TXcodebooksize$ as $\cF = \{\sfv_{1},\sfv_{2},\ldots,\sfv_{\TXcodebooksize}\}$. We similarly denote the user analog codebook with size $\RXcodebooksize$ as $\usercodebook= \{\sfg_{1},\sfg_{2},\ldots,\sfg_{\RXcodebooksize}\}$.  To obtain the analog precoder $\FRF[\horizonindex]$ and the analog combiner $\userWRF[\horizonindex]$, the base station and the user exhaustively sweep RF beams over $\TXcodebooksize\ \RXcodebooksize$ time slots simultaneously for all RF chains. 

The analog beam management procedure is performed by exchanging SS bursts, where a single SS burst comprises multiple SS blocks \cite{GioPolZor:Beam_Management_3GPP_NR_mmWave_Tutorial:COMM18}. We denote $\NSS$ as the number of SS blocks per burst and $\SSperiodicity$ as the periodicity between two SS burst exchangements. The total number of beams, $\TXcodebooksize\RXcodebooksize$, is divided into bursts of size $\NSS$ that are exchanged every $\SSperiodicity$ time slots such that the overhead of the analog beam training procedure is \cite{KimCasHea:Joint-relay-selection-beam-management:TVT23}
\begin{IEEEeqnarray}{lCr}
	\AnalogBTOverhead = \SSperiodicity\left\lceil \frac{\TXcodebooksize\ \RXcodebooksize}{\NSS} \right\rceil.
	\label{eq:BTlen_mmWave_analog_beam_training}
\end{IEEEeqnarray}
During the analog beam training, where the transmit and receive beam pair are being swept simultaneously for RF chains, the user feeds back the spectral efficiency for each transmit and receive beam pair to the base station. We use the MMSE estimator for the effective channel, which accounts for the measurement error in its estimation, under a rectangular Doppler spectrum as outlined in \cite[Sec. 4.8]{HeaLoz:MIMO-book:18}. The MMSE estimator is expressed in terms of the ratio of pilots per symbol transmission, which we denote as $\analogPilotRatio$, and the total number of OFDM frames during the analog beam training, which we denote as $\analogOFDMFrames$. Then, the MMSE can be expressed as 
\begin{IEEEeqnarray}{lCr}
	\text{MMSE}=\frac{1}{1+\analogPilotRatio\analogOFDMFrames\text{SNR}},
	\label{eq:MMSE_value}
\end{IEEEeqnarray}
and the effective SNR as 
\begin{IEEEeqnarray}{lCr}
	\text{SNR}_{\text{eff}}=\frac{\text{SNR}(1-\text{MMSE})}{1+\text{SNR}\cdot\text{MMSE}}.
	\label{eq:effective_SNR}
\end{IEEEeqnarray}
The effective SNR is applied to the spectral efficiency feedback from the user to the base station 
\begin{IEEEeqnarray}{lCr}
	\userRFSEFeedback[\horizonindex;\sfg[\horizonindex],\sfv[\horizonindex]]  = \frac{1}{\subcarriernum}\sum_{\subcarrierindex=1}^{\subcarriernum}\log _{2}\bigg(1+\text{SNR}_{\text{eff}} \bigg|\sfg^{*}[\horizonindex] \channelmatrix[\subcarrierindex,\horizonindex]\sfv[\horizonindex]\bigg|^{2}\bigg).
	\label{eq:SE_feedback}
\end{IEEEeqnarray}
We presume a greedy approach to configure the analog beamformers $\FRF[\horizonindex]$ and $\userWRF[\horizonindex]$. For RF chain pair $(\TXindexRF,\RXindexRF)$ ranging from $\TXindexRF=1,\ldots,\NTXRF$ and $\RXindexRF=1,\ldots,\userNRF$, the system obtains $\{\sfg_{\TXindexRF}^{\star}[\horizonindex], \sfv_{\RXindexRF}^{\star}[\horizonindex]\}$ by solving 
\begin{subequations}\label{eq:analog-beam-training-mmwave}
\begin{IEEEeqnarray}{s,lCr}\label{eq:objective-analog-beam-training}
$\{\sfg_{\TXindexRF}^{\star}[\horizonindex], \sfv_{\RXindexRF}^{\star}[\horizonindex]\}=$&\underset{\forall \sfg \in \mathcal{W}, \forall \sfv \in \mathcal{F}}{\max }\userRFSEFeedback[\horizonindex;\sfg[\horizonindex],\sfv[\horizonindex]]\\
subject to &\sfg\neq \sfg_{1},\ldots,\sfg\neq\sfg_{\TXindexRF-1},\\
&\sfv\neq \sfv_{1},\ldots,\sfv\neq \sfv_{\TXindexRF-1}.
\end{IEEEeqnarray}
\end{subequations}
The system then sets $\FRF[\horizonindex] = [\sfv^{\star}_{1}[\horizonindex],\sfv^{\star}_{2}[\horizonindex],\ldots,\sfv^{\star}_{\NBSRF}[\horizonindex]]$ and $\userWRF[\horizonindex] = [\sfg^{\star}_{1}[\horizonindex],\sfg^{\star}_{2}[\horizonindex],\ldots,$ $\sfg^{\star}_{\userNRF}[\horizonindex]]$. Note that the constraints $\sfg\neq \sfg_{1},\ldots,\sfg\neq \sfg_{\TXindexRF-1}$ and 
$\sfv\neq \sfv_{1},\ldots,\sfv\neq \sfv_{\TXindexRF-1}$ ensure that distinct beams are used for separate RF chains, thereby achieving spatial multiplexing gain \cite{SunQiLi:Beam-training-allocation-MU-mmWave-massive:TCOMM19}.

After analog training, the system finds the digital precoder and combiner by estimating the digital effective channel $\userEffectiveChannel[\subcarrierindex,\horizonindex]$, for all subcarriers $\subcarrierindex=1,\ldots,\subcarriernum$. We choose to represent the measurement error from pilot-based estimation using the mean squared error (MSE) \cite[Sec. 3.7]{HeaLoz:MIMO-book:18}
\begin{IEEEeqnarray}{lCr}
	\text{MSE}=\frac{1}{\frac{\digitalPilotRatio\digitalOFDMFrames}{\Nbs}\text{SNR}},
	\label{eq:MSE_value}
\end{IEEEeqnarray}
where $\digitalPilotRatio$ is the ratio of pilots per symbol transmission and $\digitalOFDMFrames$ is the total number of OFDM frames in the digital training. {Let us denote $\bm{\mathsf{\delta}}[\subcarrierindex,\horizonindex]\sim\cC\cN(0,\bI)$ as a complex Gaussian random variable independent from the digital effective channel. We model the estimated effective channel using uncertainty of the form \cite{WanAuLau:MIMO-zero-forcing-channel-estimation-error:TCOMM07}}
\begin{IEEEeqnarray}{lCr}
	\userEffectiveChannel[\subcarrierindex,\horizonindex] =  \WRF^{*}[\horizonindex]\userchannel[\subcarrierindex,\horizonindex]\FRF[\horizonindex] + \frac{1}{\sqrt{\frac{\digitalPilotRatio\digitalOFDMFrames}{\Nbs}\text{SNR}}}\bm{\mathsf{\delta}}[\subcarrierindex,\horizonindex].
	\label{eq:effective_channel}
\end{IEEEeqnarray}
With the effective channel $\userEffectiveChannel[\subcarrierindex,\horizonindex]$, the user can compute the least squares digital combiner as 
\begin{IEEEeqnarray}{lCr}
	\userWBB[\subcarrierindex,\horizonindex] = \userEffectiveChannel[\subcarrierindex,\horizonindex](\userEffectiveChannel^{*}[\subcarrierindex,\horizonindex]\userEffectiveChannel[\subcarrierindex,\horizonindex])^{-1}.
	\label{eq:digital_combiner}
\end{IEEEeqnarray}
We further assume the effective channel is quantized with RVQ codebook, denoted as $\mathcal{H}$, constructed with Lloyd's algorithm, following \cite{AlkLeuHea:Limited-feedback-hybrid-precoding-MU-mmWave:TCOMM15}. We note the PMI codebook that is used for sub-6 GHz is not currently implemented in mmWave. RVQ codebooks can be randomly generated independently from the channel realization, and are known to be asymptotically optimal regarding the number of transmit antennas and codebook size  \cite{BroLov:MIMO-nulforming-RVQ:TCOMM14}. Then, the quantized effective channel fed back from the user to the base station can be written as
\begin{IEEEeqnarray}{lCr}
\userQuantizedChannel[\subcarrierindex,\horizonindex] = \underset{\tilde{\channelmatrix}\in\mathcal{H}}{\argmax}\|\userEffectiveChannel^{*}[\subcarrierindex,\horizonindex]\tilde{\channelmatrix}[\subcarrierindex,\horizonindex]\|_{2}.
\label{eq:effective_channel_feedback}
\end{IEEEeqnarray} 
Finally, the base station computes the MMSE digital precoder as 
\begin{IEEEeqnarray}{lCr}
\userFBB[\subcarrierindex,\horizonindex] = \userQuantizedChannel^{*}[\subcarrierindex,\horizonindex](\userQuantizedChannel[\subcarrierindex,\horizonindex]\userQuantizedChannel^{*}[\subcarrierindex,\horizonindex])^{-1}.
\label{eq:digital_precoder}
\end{IEEEeqnarray}
The overhead of digital beam training, which we denote as $\DigitalBTOverhead$, only involves the feedback \eqref{eq:effective_channel_feedback} and the matrix manipulations throughout \eqref{eq:digital_combiner}, \eqref{eq:effective_channel}, and \eqref{eq:digital_precoder}. The feedback involves channel access unlike matrix manipulations, hence, the dominant factor in $\DigitalBTOverhead$ is the number of quantization bits of $\cH$. Let us denote $\RVQbits$ as the number of quantization bits of $\cH$ and $\channelbits$ as the number of bits that can be sent through the feedback channel over a single time slot. Then, the overhead of the digital beam training procedure can be written as
\begin{IEEEeqnarray}{lCr}
	\DigitalBTOverhead = \left\lceil\frac{\RVQbits}{\channelbits}\right\rceil.
	\label{eq:BTlen_mmWave_digital_beam_training}
\end{IEEEeqnarray}
Compared to the analog beam training overhead, $\DigitalBTOverhead\ll\AnalogBTOverhead$ because digital beam training requires far fewer feedback procedures than the multiple SS burst exchanges required in analog beam training. The gap between $\DigitalBTOverhead$ and $\AnalogBTOverhead$ will increase when the number of antennas in the system increases.

%===========================================================%
\subsection{Sub-6 GHz system model}\label{sec:sub6_system_model}
%===========================================================%

In the sub-6 GHz band, we assume the system employs the fully digital beamforming architecture. The base station is equipped with $\cmWaveNt$ antennas and RF chains to send $\cmWaveNS$ data streams. The user is equipped with $\cmWaveuserNr$ antennas and RF chains. The size of symbol vector $\usersymbol[\subcarrierindex,\horizonindex]$ is $\cmWaveuserNS \times 1$. The symbol vector is assumed to be normalized such that $\bbE[|\mathsf{\usersymbol}[\subcarrierindex,\horizonindex]|^{2}] = 1$. The base station precodes the symbol vector with {an} $\cmWaveNt \times \cmWaveNS$ frequency-selective precoder $\cmWaveuserFBB[\subcarrierindex, \horizonindex]$. The precoded signal propagates through {the channel denoted as} $\cmWavechannelmatrix[\subcarrierindex,\horizonindex]$ {and the noise denoted as $\cmWaveusernoise[\subcarrierindex,\horizonindex]$. We assume the noise is IID following the distribution $\cN_{C}(0,\cmWavenoiseSTD^{2})$} At the user, we assume the received signal is decoded with $\cmWaveuserNr \times \cmWaveNS$ frequency-selective decoder $\cmWaveuserWBB[\subcarrierindex,\horizonindex]$. {We set power constraint on the base station by denoting $\cmWaveuserTransmitpower$ as the transmit power and as $\|\cmWaveuserFBB[\subcarrierindex,\horizonindex]\|_{F}^{2}=\cmWaveNS$.} Then, the end-to-end input-to-output relation in the sub-6 GHz band is
\begin{equation}
\underline{\bm{\mathsf{y}}}[\subcarrierindex, \horizonindex] =\sqrt{\cmWaveuserTransmitpower\cmWaveuserlargescalefading} \cmWaveuserWBB^{*}[\subcarrierindex, \horizonindex]  \cmWaveuserchannel[\subcarrierindex,\horizonindex]   \cmWaveuserFBB[\subcarrierindex, \horizonindex]  \usersymbol[\subcarrierindex,\horizonindex]+\cmWaveuserWBB^{*}[\subcarrierindex, \horizonindex] \cmWaveusernoise[\subcarrierindex,\horizonindex],
\end{equation}
and the spectral efficiency {per the subcarrier $\subcarrierindex$} in the sub-6 GHz band can be written as
\begin{align}
\cmWaveuserSE[\subcarrierindex,\horizonindex]&= \log\det \left(\mathbf{I}_{\cmWaveuserNS}+\cmWaveuserTransmitpower\cmWaveuserlargescalefading\noiseSTD^{-2} \cmWaveuserWBB^{*}[\subcarrierindex, \horizonindex]\right.\nonumber\\&\left.\times \cmWaveuserchannel[\subcarrierindex, \horizonindex]  \cmWaveuserFBB[\subcarrierindex, \horizonindex]\cmWaveuserFBB^{*}[\horizonindex]  \cmWaveuserchannel^{*}[\subcarrierindex, \horizonindex]  \cmWaveuserWBB[\subcarrierindex, \horizonindex] \right).
\end{align}
{Due to the normalization of the symbol vector, the SNR prior to beamforming in the sub-6 GHz band is $\cmWaveuserTransmitpower\cmWaveuserlargescalefading\cmWavenoiseSTD^{-2}$. }

%===========================================================%
\subsection{Sub-6 GHz beam management procedure}\label{sec:sub6_beam_management}
%===========================================================%

The 5G NR beam management procedure in the sub-6 GHz band uses channel state information reference signals (CSI-RSs), a technique inherited from 4G, and precoding matrix indicator (PMI) feedback. We presume that the Type-1 PMI codebook is employed and the PMI feedback indicates the PMI table index, which includes both candidate precoders and the channel quantization \cite{3GPP-TS-38.214}. {Let us denote $\cmWavePilotRatio$ as the ratio of pilots per symbol transmission, $\cmWaveOFDMFrames$ as the total number of OFDM frames in the sub-6 GHz band beam training, and $\bm{\mathsf{\delta}}[\subcarrierindex,\horizonindex]\sim\cC\cN(0,\bI)$ as a complex Gaussian random variable independent from the channel $\cmWaveuserchannel[\subcarrierindex,\horizonindex]$. We model the CSI before quantization with the error model \cite{WanAuLau:MIMO-zero-forcing-channel-estimation-error:TCOMM07}}
\begin{IEEEeqnarray}{lCr}
	\precoderInfo[\subcarrierindex,\horizonindex] = \cmWaveuserchannel[\subcarrierindex,\horizonindex] + \frac{1}{\sqrt{\frac{\cmWavePilotRatio\cdot\cmWaveOFDMFrames}{\cmWaveNt}\text{SNR}}}\bm{\mathsf{\delta}}[\subcarrierindex,\horizonindex].
	\label{eq:sub6_CSI}
\end{IEEEeqnarray}
The precoder selection based on the PMI feedback can be written as
\begin{IEEEeqnarray}{lCr}
	\cmWaveuserFBB[\subcarrierindex,\horizonindex] =\underset{\forall \bm{\mathsf{F}} \in \underline{\cH}}{\argmax }\left\|\precoderInfo[\subcarrierindex,\horizonindex] \bm{\mathsf{F}}[\subcarrierindex,\horizonindex]\right\|
	\label{eq:sub6_precoder}
\end{IEEEeqnarray}
where $\underline{\cH}$ denotes the PMI codebook. We further assume the PMI feedback includes the spectral efficiency feedback
\begin{align}
	\cmWaveuserRFSEFeedback[\horizonindex] &= \frac{1}{\cmWavesubcarriernum}\sum_{\subcarrierindex=1}^{\cmWavesubcarriernum}\log _{2}\bigg(1+\text{SNR} \bigg|\cmWaveuserWBB^{*}[\subcarrierindex,\horizonindex] \precoderInfo[\subcarrierindex,\horizonindex]\cmWaveuserFBB[\subcarrierindex,\horizonindex]\bigg|^{2}\bigg),
	\label{eq:sub6_SE_feedback}
\end{align}
where the combiner $\cmWaveuserWBB^{*}[\subcarrierindex,\horizonindex]$ is computed based on zero-forcing. {Let us denote $\PMIcodebooksize$ as the size of the PMI codebook and $\cmWavechannelbits$ as the number of bits that can be sent through the sub-6 GHz feedback channel over a single time slot. The overhead from the beam training procedure in the sub-6 GHz band, which we denote as $\cmWaveBTOverhead$, can be written as
\begin{IEEEeqnarray}{lCr}
	\cmWaveBTOverhead = \left\lceil\frac{\log_{2}\PMIcodebooksize}{\cmWavechannelbits}\right\rceil.
	\label{eq:BTlen_sub6_digital_beam_training}
\end{IEEEeqnarray}}
Typically, the beam training overhead in the sub-6 GHz band is between the analog beam training overhead and digital beam training overhead in the mmWave band such that $\DigitalBTOverhead<\cmWaveBTOverhead<\AnalogBTOverhead$. It is noteworthy that the beam training overhead in the sub-6 GHz can be reduced by grouping multiple antennas in the base station to a single port to reduce the beam training overhead \cite{LeeKimCho:DL-channel-reconstruction-massive-MIMO:TCOMM21}.

%%%%%%%%%%%%%%%%%%%%%%%%%%%%%%%%%%%%%%%%%%%%%%%%%%%%%%%%%%%%%
\section{Reinforcement learning formulation of the joint band assignment and beam management problem}\label{sec:problem_definition}
%%%%%%%%%%%%%%%%%%%%%%%%%%%%%%%%%%%%%%%%%%%%%%%%%%%%%%%%%%%%%

In this section, we formulate the joint band assignment and beam management problem for the multi-band wireless network as an RL problem. We first describe the underlying learning model as an MDP. We then discuss the challenges of the RL formulation by describing the nonstationarity of the action space over different bands. {We describe a baseline approach of the RL formulation using action masking in \secref{sec:action_masking_baseline}.} We further detail the remedies on the challenges by proposing an algorithm based on HRL in \secref{sec:main_algorithm}.

The base station aims to maximize the system's data rate by selecting the best band of operation and precoder at each time slot. {For each time slot $\horizonindex$, we denote the actions that the transmitter can take as $\action[\horizonindex]$. The action dictates a chosen band and also whether to perform beam training or data transmission. We say the action is a set including a chosen band $\bandassign[\horizonindex]$ and a beam management mode $\modeindex[\horizonindex]$.} Specifically, we set $\bandassign[\horizonindex]=1$ to imply the mmWave band being the band of operation and $\bandassign[\horizonindex]=0$ to imply the sub-6 GHz band being the band of operation. We also set $\modeindex[\horizonindex]=1$ to indicate data transmission and $\modeindex[\horizonindex]=0$ to indicate beam training. The system's data rate, which is the performance metric of interest, can be written as
\begin{align}
\userRate[\horizonindex] &= \left((1-\bandassign[\horizonindex])\frac{\cmWavebandwidth}{\cmWavesubcarriernum}\sum_{\subcarrierindex=1}^{\cmWavesubcarriernum} \cmWaveuserSE[\subcarrierindex,\horizonindex]+\bandassign[\horizonindex]\frac{\mmWavebandwidth}{\mmWavesubcarriernum}\sum_{\subcarrierindex=1}^{\mmWavesubcarriernum}\userSE[\subcarrierindex,\horizonindex]\right).
\label{eq:total-rate}
\end{align}
We assume that $\horizonlen$ is finite to keep the cumulative data rate finite. Denoting the binary variable $c(\action[\horizonindex])=0$ when beam training is in progress and $c(\action[\horizonindex])=1$ when data transmission is performed, the optimization problem can be written as
\begin{equation}
\max_{\{\action[\horizonindex]\}} \sum_{\horizonindex=1}^{\horizonlen} c(\action[\horizonindex])\userRate[\horizonindex].
\label{eq:ch2-optimization}
\end{equation}

We solve \eqref{eq:ch2-optimization} by formulating an MDP, which has been shown to be an effective approach for many resource allocation problems \cite{LuoHoaGon:DRL_application_Communications_Survey:COMM19}. {The crucial elements in an MDP are the state space, action space, and the reward function. The design of the action space poses a significant challenge in the MDP formulation, as the choice between discrete and continuous action spaces plays a crucial role in determining scalability. On the one hand, discrete action spaces may lead to scalability issues due to the need for defining separate actions for various combinations of bands and beam management modes. On the other hand, continuous action spaces offer a partial solution to scalability concerns by replacing discrete actions with continuous variables that act as decision boundaries \cite{KimCasHea:Joint-relay-selection-beam-management:TVT23}. We first describe the MDP of the DRL approach using continuous action space in \secref{sec:action_masking_baseline}. The distinct beam management procedures in the mmWave band and sub-6 GHz band pose limitations on the performance of continuous RL algorithms, as decision boundaries are applied uniformly across all bands. In this regard, we further specify the MDP of the HRL-based algorithm in \secref{sec:main_algorithm}.}

%%%%%%%%%%%%%%%%%%%%%%%%%%%%%%%%%%%%%%%%%%%%%%%%%%%%%%%%%%%%%
\section{Reinforcement learning approach of the joint band assignment and beam management problem using action masking}\label{sec:action_masking_baseline}
%%%%%%%%%%%%%%%%%%%%%%%%%%%%%%%%%%%%%%%%%%%%%%%%%%%%%%%%%%%%%

In this section, we present a DRL-based method for solving the joint band assignment and beam management problem using threshold-based actions. The DRL-based approach serves as a baseline to the HRL-based algorithm and an example application of continuous action spaces with action masking, later used in the HRL-based algorithm as well.

A naive approach to addressing inconsistent action spaces is through action masking, where the total action space includes all possible actions conservatively, and invalid action probabilities are forced to zero  \cite{KanSchHau:Action-space-shaping-DRL:COG20,HuaOnt:Invalid-action-masking-policy-gradient:FLAIRS22}.  We describe the MDP formulation of the brute-force approach using action space masking, which we later set as a baseline in the experiments under the name three-threshold policy. 
The state $\state[\horizonindex]$, action $ \action[\horizonindex]$, and reward $r[\horizonindex]$ of the three-threshold policy can be described as the following.

\textit{1) States:} The state space incorporates the selected beamformers and feedback used throughout the beam management procedures as discussed in \secref{sec:beam_management} and \secref{sec:sub6_system_model}. In the mmWave band, the analog beam training determined the analog beamformers based on spectral efficiency feedback followed by digital effective channel estimation. In the sub-6 GHz band, the PMI feedback determines the precoder computation. The state can be written as  
\begin{IEEEeqnarray}{lCr}
\state[\horizonindex] = \Bigl\{\FRF[\horizonindex],\userWRF[\horizonindex], \userRFSEFeedback[\horizonindex], \bigl\{\userQuantizedChannel[\subcarrierindex,\horizonindex]\bigr\}^{\subcarriernum}_{\subcarrierindex=1},\nonumber\\ \bigl\{\cmWaveuserFBB[\subcarrierindex,\horizonindex]\bigr\}^{\cmWavesubcarriernum}_{\subcarrierindex=1},\bigl\{\precoderInfo[\subcarrierindex,\horizonindex]\bigr\}^{\cmWavesubcarriernum}_{\subcarrierindex=1}\Bigr\}.
\label{eq:DRL_state}
\end{IEEEeqnarray}
Note that the codebook assumption for constructing the analog beamformers $\FRF[\horizonindex],\userWRF[\horizonindex]$, the quantized feedback channel $\bigl\{\userQuantizedChannel[\subcarrierindex,\horizonindex]\bigr\}^{\subcarriernum}_{\subcarrierindex=1}$ in the mmWave band, and the precoder $\bigl\{\cmWaveuserFBB[\subcarrierindex,\horizonindex]\bigr\}^{\cmWavesubcarriernum}_{\subcarrierindex=1}$ in sub-6 GHz band can be used to reduce the state space dimension.

\textit{2) Actions:} The action space consist of three continuous variables
\begin{align}
\action[\horizonindex] = \{\underline{\tau}[\horizonindex],\DRLanalogAction[\horizonindex],\DRLdigitalAction[\horizonindex]\}.
\label{eq:DRL_action}
\end{align} 
{The spectral efficiency feedback of the operating band is compared with the thresholds. At mmWave, the} spectral efficiency feedback $\userRFSEFeedback[\horizonindex]$ is compared with each threshold to perform one of the following. When $\userRFSEFeedback[\horizonindex]<\underline{\tau}[\horizonindex]$, the base station switches band to the sub-6 GHz. When $\underline{\tau}[\horizonindex]<\userRFSEFeedback[\horizonindex]<\DRLanalogAction[\horizonindex]$, the base station tries analog beam training. When $\DRLanalogAction[\horizonindex]<\userRFSEFeedback[\horizonindex]<\DRLdigitalAction[\horizonindex]$, the base station triggers digital beam training. When $\DRLdigitalAction[\horizonindex]<\userRFSEFeedback[\horizonindex]$, the base station keeps both the analog and digital precoders and transmits data. In the sub-6 GHz band, the threshold $\DRLanalogAction[\horizonindex]$ is masked to compare the spectral efficiency to determine band switching and transmission mode. When $\cmWaveuserRFSEFeedback[\horizonindex]<\underline{\tau}[\horizonindex]$, the base station switches band to the mmWave. When $\underline{\tau}[\horizonindex]<\cmWaveuserRFSEFeedback[\horizonindex]<\DRLdigitalAction[\horizonindex]$, the base station tries beam training. When $\DRLdigitalAction[\horizonindex]<\cmWaveuserRFSEFeedback[\horizonindex]$, the base station keeps the precoder and transmits data. 

\textit{3) Reward:} The reward can be written as
\begin{align}
r(\state[\horizonindex],\action[\horizonindex]) = c(\action[\horizonindex])\userRate[\horizonindex],
\label{eq:DRL_reward}
\end{align} 
since the objective of an MDP is maximizing the cumulative reward, as in \eqref{eq:ch2-optimization}, over time.

The main issue with the MDP specified by \eqref{eq:DRL_state}, \eqref{eq:DRL_action}, and \eqref{eq:DRL_reward} lies in the design of the action space. Though an implicit assumption made in the MDP formulations of \cite{LuoHoaGon:DRL_application_Communications_Survey:COMM19} and references therein is that the state space and action space are invariant over time, the joint band assignment and beam management problem \eqref{eq:ch2-optimization} introduces the dependence of the beam management procedure on the operating band. This distinctive beam training procedures in each band lead to a inconsistent action space making it difficult to formulate an MDP. Rather, the band selection and beam training decisions need to be learned individually to keep the action space consistent within a single policy. In the following section, we describe how hierarchical learning can be used to solve these issues.

%%%%%%%%%%%%%%%%%%%%%%%%%%%%%%%%%%%%%%%%%%%%%%%%%%%%%%%%%%%%%
\section{Hierarchical reinforcement learning algorithm for joint band assignment and beam management } \label{sec:main_algorithm}
%%%%%%%%%%%%%%%%%%%%%%%%%%%%%%%%%%%%%%%%%%%%%%%%%%%%%%%%%%%%%

In this section, we propose an HRL-based algorithm for solving the joint band assignment and beam management problem. We first give a brief introduction of HRL in \secref{sec:HRL}, focusing on the relation of state, action, and reward defined in the upper-level and lower-level policies. We then describe the proposed algorithm, a novel approach for the joint band assignment and beam management problem, incorporating  off-policy correction methods and an adaptive upper-level policy period in \secref{sec:learning_algorithm}.

%===========================================================%
\subsection{Hierarchical reinforcement learning}\label{sec:HRL}
%===========================================================%

\begin{figure}[t]
	\centering
	\includegraphics[width=0.6\columnwidth,draft=false]{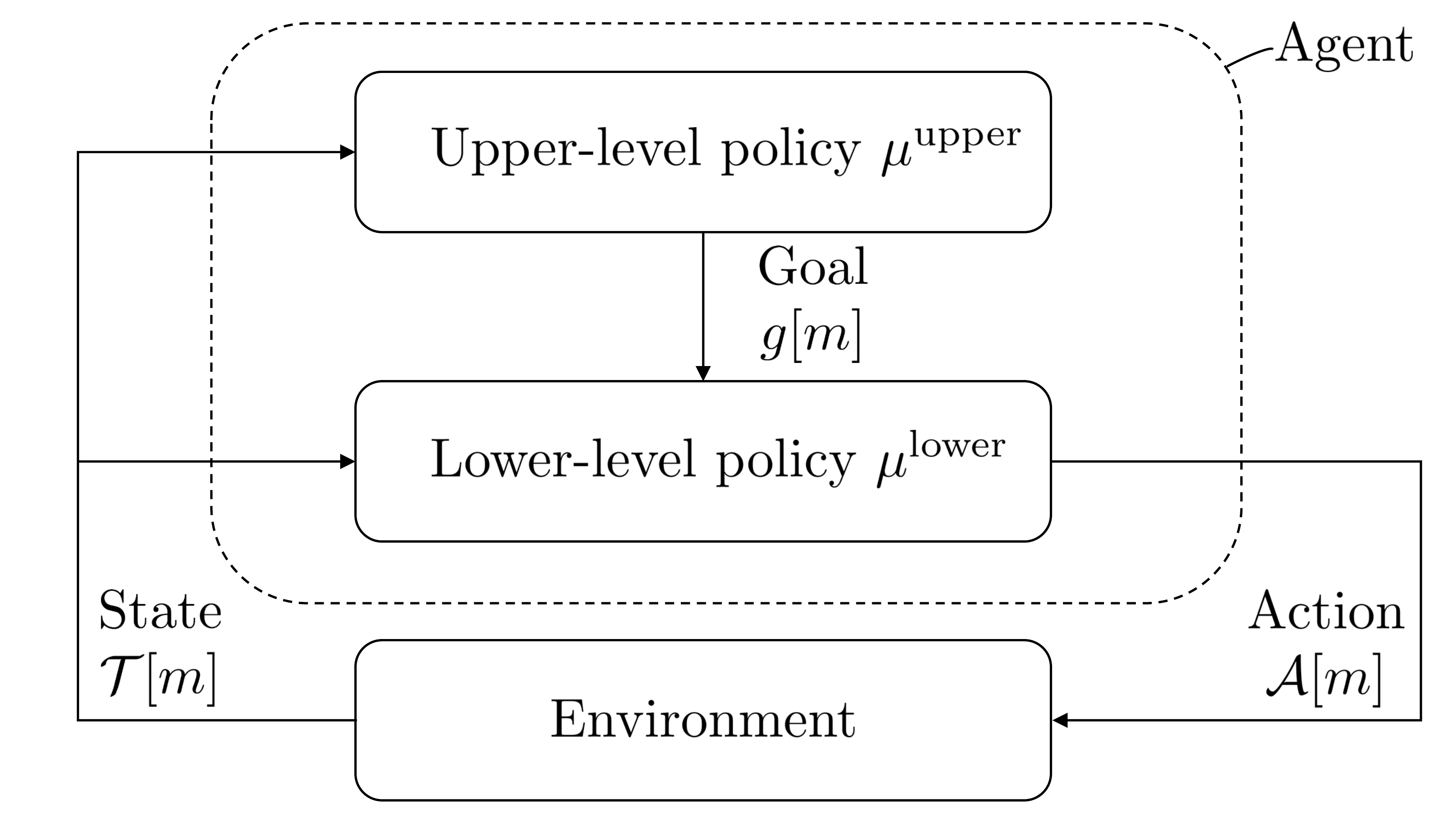}
	\caption{Hierarchy between the upper-level and lower-level policy in the HRL framework. The upper-level policy generates goals as its action, which is inputted to the lower-level policy to determine the action interacting with the environment. }
	\label{fig:HRL}
\end{figure}

HRL algorithms build upon RL algorithms, which aim to find the policy that maximizes the cumulative reward by training neural networks. The key difference of HRL algorithms to traditional RL algorithms lies in the separation of decision layers, which represents the decomposition of the complex task given to the decision-making agent. The upper decision layer selects subtasks to be performed and the lower decision layer executes the chosen subtask. In the RL framework, the policy of the agent maps a state $\state$ to an action $\action$. HRL algorithms, depicted in \figref{fig:HRL}, extend the framework to consist the upper-level policy $\upperPolicy$ and the lower-level policy $\lowerPolicy$ \cite{NacGuLev:Data-efficient-HRL:NIPS18}. The upper-level policy maps a state to a high-level action (or \emph{goal}), where the lower-level policy maps a pair $(\state,\upperAction)$ to an action $\action$.

%We provide a general overview of DDPG described on a generic policy $\mu$, where $\mu$ can be substituted by the upper-level policy $\upperPolicy$ and the lower-level policy $\lowerPolicy$. 

We use DDPG~\cite{LilHunPri:DDPG:15} to train the upper-level policy $\upperPolicy$ and the lower-level policy $\lowerPolicy$. Four neural networks are trained in DDPG, where each neural network {corresponds} to the online actor network $\thetaA$, the target actor network $\targthetaA$, the online critic network $\thetaC$, and the target critic network $\targthetaC$. The actor networks represent a policy, whereas the critic networks evaluate a policy. The target networks are delayed copies of the online networks with slow updates, which helps to reduce the effects of overfitting and instability.

DDPG uses experience replay that stores a buffer of experiences to update the neural networks. The experience replay consist of trajectories, where a single trajectory is a tuple of the state, action, reward and successor state. The trajectory of the lower-level policy is a tuple of $(\state[\horizonindex],\upperAction[\horizonindex],\action[\horizonindex],\intrinsicReward[\horizonindex],\state[\horizonindex+1])$, where $\intrinsicReward$ is the intrinsic reward provided by the upper-level policy. The update of the neural networks for the lower-level policy incorporates the goals in the typical loss minimization and policy gradient methods. Specifically, a $\batchsize$-element randomly sampled minibatch is from the experience replay of the lower-level policy, which we denote as $\lowerExperienceReplay$. Using the minibatch, the lower-level $\thetaC$ is updated by minimizing the loss 
\begin{align}
L &= \frac{1}{\batchsize}\sum_{\horizonindex'}\bigg((\intrinsicReward[\horizonindex'] + \gamma \lowerQ(\state[\horizonindex'+1,\upperAction[\horizonindex'+1]],\mu_{\targthetaA}(\state[\horizonindex'+1],\upperAction[\horizonindex'+1])|\targthetaC) \nonumber\\ &- \lowerQ(\state[\horizonindex'],\upperAction[\horizonindex'],\action[\horizonindex']|\thetaC) )^{2} \bigg).
\label{eq:lower_level_DDPG_loss_function}
\end{align}	
The lower-level $\thetaA$ is updated with the policy gradient 
 \begin{align}
 &\sum_{\horizonindex'}\frac{1}{\batchsize}\bigg(\nabla_{\action}\lowerQ(\state,\upperAction,\action|\thetaC)|_{\state=\state[\horizonindex'],\upperAction=\upperAction[\horizonindex'],\action=\mu_{\thetaA}(\state[\horizonindex'],\upperAction[\horizonindex'])}\\&\times
 \nabla_{\thetaA}\mu_{\thetaA}(\state,\upperAction)|_{\state=\state[\horizonindex'],\upperAction = ,\upperAction[\horizonindex']}\bigg). \label{eq:lower_level_DDPG_actor_network_sampled_policy_gradient}
 \end{align}
The target networks are slowly updated from the online networks, where the parameter $\eta<<1$ controls the variance of the target networks:
\begin{IEEEeqnarray}{lCr}
	\targthetaA &\leftarrow \eta\thetaA+(1-\eta)\targthetaA, \nonumber\\
	\targthetaC &\leftarrow \eta\thetaC+(1-\eta)\targthetaC.
	\label{eq:DDPG_target_from_online}
\end{IEEEeqnarray}
The parameter $\eta$ can help alleviate the overestimation of the Q-values \cite{FujHooMeg:TD3:ICML18}.

The upper-level trajectory involves multiple time steps, which we denote as $\upperPeriod$, because the subtask sampling happens coarsely. A single upper-level transition is a tuple of 
\begin{IEEEeqnarray}{lCr}
\bigg(\bigl\{\state[\horizonindex']\bigr\}^{\horizonindex+\upperPeriod-1}_{\horizonindex'=\horizonindex},\bigl\{\upperAction[\horizonindex']\bigr\}^{\horizonindex+\upperPeriod-1}_{\horizonindex'=\horizonindex}\nonumber\\,\bigl\{\action[\horizonindex']\bigr\}^{\horizonindex+\upperPeriod-1}_{\horizonindex'=\horizonindex}
,\bigl\{\envReward[\horizonindex']\bigr\}^{\horizonindex+\upperPeriod}_{\horizonindex'=\horizonindex},\state[\horizonindex+\upperPeriod]\bigg),
\label{eq:upper-level-trajectory}
\end{IEEEeqnarray}
where $\envReward$ is the reward given by the environment. To describe the usage practical actor-critic algorithms such as DDPG based on $\upperExperienceReplay$, we denote the aggregated state as $\aggUpperState[\horizonindex] = (\state[\horizonindex],\ldots,\state[\horizonindex+\upperPeriod-1])$, aggregated goal as $\aggGoal[\horizonindex] = (\upperAction[\horizonindex],\ldots,\upperAction[\horizonindex+\upperPeriod-1])$, and cumulative environmental reward as $\aggUpperReward =  \sum_{\horizonindex'=\horizonindex}^{\horizonindex+\upperPeriod}\envReward[\horizonindex']$.

%In terms of the loss minimization and policy gradient methods as in \eqref{eq:lower_level_DDPG_loss_function} and \eqref{eq:lower_level_DDPG_actor_network_sampled_policy_gradient}, the state, goal, action, and reward can be substituted by the aggregated state $(\state[\horizonindex],\ldots,\state[\horizonindex+\upperPeriod-1])$, goal $(\upperAction[\horizonindex],\ldots,\upperAction[\horizonindex+\upperPeriod-1])$, and the cumulative reward $\sum_{\horizonindex'=\horizonindex}^{\horizonindex+\upperPeriod}\envReward[\horizonindex']$.  

%, action $(\action[\horizonindex]\ldots,\action[\horizonindex+\upperPeriod-1])$

When updating the upper-level $\thetaC$, minimizing the loss as in \eqref{eq:lower_level_DDPG_loss_function}, an off-policy correction is required to address the varying $\lowerPolicy$ in a single upper-level trajectory. {Let us denote $\lowerbaselinePolicy$ as the lower-level policy that was used when sampling $\upperExperienceReplay$ and $\lowerPolicy$ as the current lower-level policy. We use importance sampling to estimate the loss function regarding the samples generated with $\lowerPolicy$ based on the experience replay $\upperExperienceReplay$ from lower-level policy following $\lowerbaselinePolicy$.} The direct importance correction can be written as \cite{NacGuLev:Data-efficient-HRL:NIPS18}
\begin{IEEEeqnarray}{lCr}
	\importanceWeight = \prod_{\horizonindex'=\horizonindex}^{\horizonindex+\upperPeriod-1}\frac{\lowerPolicy(\state[\horizonindex'],\upperAction[\horizonindex'],\action[\horizonindex'])}{\lowerbaselinePolicy(\state[\horizonindex'],\upperAction[\horizonindex'],\action[\horizonindex'])},
	\label{eq:direct_importance_correction_weight}
\end{IEEEeqnarray}
which is applied in the update of upper-level $\thetaC$ in the loss function 
\begin{align}
	L &= \frac{1}{\batchsize}\sum_{\horizonindex'}\bigg((\envReward[\horizonindex'] + \gamma\importanceWeight \upperQ(\aggUpperState[\horizonindex'+1],\mu_{\targthetaA}(\aggUpperState[\horizonindex'+1])|\targthetaC) \nonumber\\  &- \upperQ(\aggUpperState[\horizonindex'],\aggGoal[\horizonindex']|\thetaC) )^{2} \bigg).
	\label{eq:upper_level_direct_importance_correction_DDPG_loss_function}
\end{align}
In case of the goal, a single goal may be selected despite the aggregated goals in the upper-level trajectory. A single goal represents the subtask selection of the upper-level policy. The corrected goal can be computed by \cite{NacGuLev:Data-efficient-HRL:NIPS18}
 \begin{IEEEeqnarray}{lCr}
 	\relabedGoal[\horizonindex] = \argmin_{\upperAction[\horizonindex]}\left(1- \importanceWeight\right)^{2},
 	\label{eq:importance_action_relabling}
 \end{IEEEeqnarray}
 which is applied in the update of the upper-level $\thetaA$ with the policy gradient 
\begin{align}
&\sum_{\horizonindex'}\frac{1}{\batchsize}\bigg(\nabla_{\action}\upperQ(\aggUpperState,\aggGoal|\thetaC)|_{\aggUpperState=\aggUpperState[\horizonindex'],\aggGoal=\mu_{\thetaA}(\aggUpperState[\horizonindex'])}\nonumber\\&\times\nabla_{\thetaA}\mu_{\thetaA}(\state)|_{\aggUpperState=\aggUpperState[\horizonindex']}\bigg).
\label{eq:upper_level_DDPG_actor_network_sampled_policy_gradient}
\end{align}
Later in the experiments, we use each off-policy correction methods as baselines.

% \eqref{eq:upper_level_direct_importance_correction_DDPG_loss_function}.
% \eqref{eq:upper_level_DDPG_actor_network_sampled_policy_gradient}

%===========================================================%
\subsection{Joint band assignment and beam management strategy based on hierarchical reinforcement learning}\label{sec:learning_algorithm}
%===========================================================%

In HRL, the upper-level policy provides its action or the goal to the lower-level policy at fixed intervals $\upperPeriod$. However, previous work has shown that setting the duration too long or too short can result in performance deterioration \cite{GenLiuWan:HRL-relay-selection-power-optimization:TCOMM21}. On the one hand, if the duration is too long, the lower-level policy may not receive enough goals to be trained effectively. On the other hand, if the duration is too short, the upper-level policy may not capture sufficient abstraction from the environment regarding the beam management overhead to guide the lower-level policy.

To address this tradeoff, we propose the use of round skipping, which is inspired by bandit algorithms \cite{DicSanSri:Round-skipping-combinatorial-bandits:ACM21}. The idea is to set a short default period but periodically evaluate the interaction between the agent and the environment to determine if it is unnecessarily brief. By doing so, we can ensure that the lower-level policy receives sufficient goals for training without compromising the efficiency of task decomposition. This approach offers a more flexible and adaptive solution to the challenge of setting the upper-policy period in HRL. The round skipping probability is computed based on the mean reward and action availability. Specifically, the non-skipping probability is $\min\{1,\frac{\AnalogBTOverhead}{2\AnalogBTOverhead-1}\frac{1}{\actAvailProb}\}$, where $\actAvailProb$ is the probability that action $\action$ is available at time slot $\horizonindex$ based on the history up to time slot $\horizonindex$. 

The upper-level actor-critic update is triggered every $\upperPeriod$ time slots. If the round-skipping occurs, the band assignment variable $\bandassign$ and goal $\upperAction$ is kept constant to be used in the lower-level policy computation. Otherwise, the upper-level experience replay is generated by aggregating state, action, and cumulating the environmental reward over time horizon $\horizonindex,\ldots,\horizonindex+\upperPeriod$. In the upper-level trajectory, the length of elements are truncated to $\AnalogBTOverhead$ when $\upperPeriod>\AnalogBTOverhead$. The off-policy correction is applied to take account of the varying lower-level policy. The lower-level policy uses the goal $\upperAction$ and intrinsic reward $\intrinsicReward$ given by the upper-level actor-critic networks.

	The state $\state[\horizonindex]$, goal $\upperAction[\horizonindex]$, action $ \action[\horizonindex]$, intrinsic reward $\intrinsicReward[\horizonindex]$, and extrinsic reward $\envReward[\horizonindex]$ of the HRL-based joint band assignment and beam management algorithm can be described as the following.

	\textit{1) States:} The state space in the proposed HRL algorithm is the same as the DRL method, given as \eqref{eq:DRL_state} to include the beamformers and feedback used throughout the beam management procedures as discussed in \secref{sec:beam_management} and \secref{sec:sub6_system_model}. 

	\textit{2) Goal:} The goal corresponds to the band of operation. The goal is set to $\upperAction[\horizonindex]=1$ when the mmWave band is used. Otherwise, the goal is set to $\upperAction[\horizonindex]=0$. 

\textit{3) Action:} The action space consist of two continuous variables
	\begin{align}
	\action[\horizonindex] = \{\HRLanalogAction[\horizonindex],\HRLdigitalAction[\horizonindex]\}.
	\label{eq:HRL_action}
	\end{align} 
The action space is similar to that of the DRL baseline in \eqref{eq:DRL_action} with the difference that the band assignment is determined in the upper-level action instead. The spectral efficiency feedback $\userRFSEFeedback[\horizonindex]$ at mmWave is compared with the thresholds to perform one of the following. When $\userRFSEFeedback[\horizonindex]<\HRLanalogAction[\horizonindex]$, the base station performs analog beam training. When $\HRLanalogAction[\horizonindex]<\userRFSEFeedback[\horizonindex]<\HRLdigitalAction[\horizonindex]$, the base station proceeds digital beam training. When $\HRLdigitalAction[\horizonindex]<\userRFSEFeedback[\horizonindex]$, the base station transmits data using symbols. At sub-6 GHz, $\HRLanalogAction[\horizonindex]$ is masked. When $\cmWaveuserRFSEFeedback[\horizonindex]<\HRLdigitalAction[\horizonindex]$, the base station processes beam training. When $\HRLdigitalAction[\horizonindex]<\cmWaveuserRFSEFeedback[\horizonindex]$, the base station transmits data using symbols.

{
	\textit{4) Intrinsic reward:} The intrinsic reward for solving \eqref{eq:ch2-optimization} can be written as
	\begin{align}
	\intrinsicReward(\state[\horizonindex],\upperAction[\horizonindex],\action[\horizonindex]) = c(\action[\horizonindex])\userRate[\horizonindex].
	\label{eq:HRL_intrinsic_reward}
	\end{align} 
	Note that $\upperAction[\horizonindex]$ is analogous to the band assignment variable $\bandassign[\horizonindex]$ discussed in \secref{sec:problem_definition}. 
}

{\textit{5) Extrinsic reward:} The reward provided by the environment accounts for the upper-level policy period $\upperPeriod$ such that
	\begin{align}
	\envReward[\horizonindex] = \frac{1}{\upperPeriod}\sum_{\horizonindex'}^{\horizonindex'+\upperPeriod-1}r[\horizonindex'],
	\label{eq:HRL_extrinsic_reward}
	\end{align} 
}To ensure the algorithm implementation runs within a single OFDM time slot, we note that graphics processing unit (GPU) with high clock speed and field-programmable gate
array (FPGA) may be exploited as discussed in \cite{MaWanYan:AI-application-survey-autonomous-vehicles:20}. For completeness, the pseudocode is given in \algref{alg:HRL-policy}.

\begin{algorithm}[]
	\caption{Joint band assignment and beam management strategy based on HRL}
	\label{alg:HRL-policy}
	\begin{algorithmic}[1]
		\STATE{Input: Length $\horizonlen$ of decision horizon, Boolean constant \textit{UseActionRelabling}, Boolean random variable \textit{RoundSkip}}
		\STATE{Randomly initialize online critic network $Q(s,a|\thetaC)$ and online actor network $\mu(s|\thetaA)$ with $\thetaC$ and $\thetaA$ for upper-level and lower-level}
		\FOR{$\horizonindex=1,\ldots,\horizonlen$}
		\IF{\textit{RoundSkip}}
		\STATE{Continue using upper-level action $\upperAction[\horizonindex]$}
		\ELSE
		%\FOR{$\horizonindex' = \horizonindex,\ldots,\horizonindex+\upperPeriod$}
		\STATE{Set aggregated state as $\aggUpperState[\horizonindex] = \state[\horizonindex':\horizonindex'+\upperPeriod-1]$}
		\IF{\textit{UseActionRelabling}}
		\STATE{Set goal as \eqref{eq:importance_action_relabling} }
		\ELSE
		\STATE{Set aggregated goal as $\aggGoal = \upperAction[\horizonindex':\horizonindex'+\upperPeriod-1]$}
		\ENDIF
		\STATE{Set reward as $\sum\envReward[\horizonindex']$}
		%\ENDFOR
		\STATE{Get successor state $\state[\horizonindex+\upperPeriod]$}
		\STATE{Store transition \eqref{eq:upper-level-trajectory} in $\upperExperienceReplay$}%\STATE{// Update neural networks}
		\STATE{Sample a random minibatch of $\batchsize$ transitions  from $\lowerExperienceReplay$}
		\STATE{Update the upper-level online critic network by minimizing the loss \eqref{eq:upper_level_direct_importance_correction_DDPG_loss_function}}
		\STATE{Update the upper-level online actor network by policy gradient \eqref{eq:upper_level_DDPG_actor_network_sampled_policy_gradient}}
		\STATE{Update the target networks from the online networks according to \eqref{eq:DDPG_target_from_online}}
		\STATE{Compute upper-level action $\upperAction[\horizonindex]$ by \eqref{eq:importance_action_relabling}}
		\STATE{Update $\bandassign[\horizonindex+\upperPeriod]$}
		\ENDIF
		\STATE{Select lower-level action $\action[\horizonindex]$ according to the current online actor network and exploration noise distribution $\cN$}
		\STATE{Set reward as $\intrinsicReward[\horizonindex]$ \eqref{eq:total-rate} }
		\STATE{Update $\modeindex[\horizonindex+1]$}
		%\STATE{Update $\bandassign[\horizonindex+1]$ and $\modeindex[\horizonindex+1]$}
		\STATE{Get successor state $\state[\horizonindex+1]$}
		\STATE{Store transition $(\state[\horizonindex],\upperAction[\horizonindex],\action[\horizonindex],\intrinsicReward[\horizonindex],\state[\horizonindex+1])$ in $\cD_{\text{lower}}$}%\STATE{// Update neural networks}
		\STATE{Sample a random minibatch of $\batchsize$ transitions  from $\cD_{\text{lower}}$}
		\STATE{Update the lower-level online critic network by minimizing the loss \eqref{eq:lower_level_DDPG_loss_function}}
		\STATE{Update the lower-level online actor network by policy gradient \eqref{eq:lower_level_DDPG_actor_network_sampled_policy_gradient}}
		\STATE{Update the target networks from the online networks according to \eqref{eq:DDPG_target_from_online}}	
		\ENDFOR
	\end{algorithmic}
\end{algorithm}

%%%%%%%%%%%%%%%%%%%%%%%%%%%%%%%%%%%%%%%%%%%%%%%%%%%%%%%%%%%%%
\section{Experimental results} \label{sec:experiments}
%%%%%%%%%%%%%%%%%%%%%%%%%%%%%%%%%%%%%%%%%%%%%%%%%%%%%%%%%%%%%

In this section, we evaluate the proposed HRL algorithm on a realistic multi-band wireless network. We first describe the scenario in \secref{sec:simulation_setup}. We outline the performance metric of interest and baselines in \secref{sec:performance_metric}. We then provide the numerical results and discussion in \secref{sec:results_discussion}.

%===========================================================%
\subsection{Simulation setup} \label{sec:simulation_setup}
%===========================================================%

We simulate an urban vehicular network consisting of a static base station with a fixed transmit power in mmWave and sub-6 GHz bands and mobile vehicle nodes. We implement the Manhattan mobility model, which represents urban roads with a typical grid topology found in metropolitan cities. To generate vehicle trajectories, we employ Simulation of Urban MObility (SUMO) \cite{KraErdBeh:SUMO:12}. Among the simulated vehicles, we select a single vehicle to serve as the user. We then apply the SUMO-generated vehicle trajectory to QUAsi Deterministic RadIo channel GenerAtor (QuaDRiGa), where QuaDRiGa generates the channels accounting for the geometric consideration of vehicles acting as reflectors and blockages \cite{JaeRasThi:QuaDRiGa:TAP14}. We use the 3GPP 3D Urban micro (UMi) model provided within QuaDRiGa that determines parameters such as the path, ray, complex path gain, angle of arrival, and angle of departure. At sub-6 GHz, we use the '3gpp-3d' type of antenna array provided by QuaDRiGa in accordance with the 3GPP technical report 36.873 \cite{3GPP-TR-36.873-v12.5.0}.

We summarize the key simulation parameters, which are uniformly applied to simulations unless mentioned otherwise, and assumptions as the following:

\textit{1) Mobility parameters:} When vehicles move through a crossroad, the probability of going straight is 0.5, turning left is 0.25, and turning right is 0.25. We set the average vehicle speed as 40 km/h and the vehicle density as 10 vehicles per kilometer.

\textit{2) Array and band parameters:} We assume the number of antennas at the base station and the user are $\Nbs=32$ and $\userNr=16$ at mmWave and $\cmWaveNt=4$ and $\cmWaveuserNr=4$ at sub-6 GHz. The number of streams are $\NS=\cmWaveNS=4$ and the number of RF chain are $\NBSRF=8$ at mmWave. We assume a uniform linear array (ULA) with half-wavelength spacing used at mmWave. We assume the mmWave and sub-6 GHz arrays are co-located and aligned. Note that the aligned arrays imply that the physical line-of-sight (LOS) between the base station and the user is invariant over the sub-6 GHz and mmWave bands. We select $\mmWavesubcarriernum=256$ OFDM subcarriers at mmWave and $\cmWavesubcarriernum=32$ subcarriers at the sub-6 GHz band. The sub-6 GHz band has 150 MHz bandwidth {and the mmWave band} has 850 MHz bandwidth \cite{AliGonHea:Out-of-band-mmWave-beam-selection:TCOMM18}. 

\textit{3) Beam management parameters:} In the mmWave band, we apply beam management with $\SSperiodicity=1$ and $\NSS=4$. We assume single bit limited feedback and set $\channelbits=\cmWavechannelbits=1$. We assume that a discrete Fourier transform (DFT) codebook is employed at mmWave and the Type-I PMI codebook is used at sub-6 GHz.

% The array response vector for a $N$-element ULA is given as 
%\begin{IEEEeqnarray}{lCr}
%	\ba(\phi) = \frac{1}{\sqrt{N}}\left[1, e^{-\jj\pi \cos (\phi)}, \ldots,e^{-\jj(N-1)\pi \cos (\phi)}\right]^{\mathrm{T}},
%	\label{eq:array_response_vector}
%\end{IEEEeqnarray}
%where $\phi$ is the steering angle and $\lambda$ is the carrier wavelength.

%===========================================================%
\subsection{Performance metric and baseline policies} \label{sec:performance_metric}
%===========================================================%

We evaluate the cumulative rate as specified in \eqref{eq:total-rate}. We approximate the ensemble mean by averaging over 1,000 channel instances generated by SUMO and QuaDRiGa. For the performance of the learning-based policy, either DRL-based or HRL-based, we measure the average of the last 20 iterations out of the $\horizonlen=200$ total iterations to represent the converged reward. 

We compare the proposed HRL-based algorithm to three baseline policies:
\begin{itemize}
	\item \textbf{Genie-aided} policy: This algorithm has perfect knowledge of the channel on both the mmWave and sub-6 GHz bands. 	Subsequently, this policy chooses the data transmission action with the correct frequency band and the best beam indices. Thus, the performance achieved by the genie-aided policy represents the theoretical upper limit of the system.
	\item \textbf{Three-threshold} policy: This algorithm applies DRL using threshold-based actions. The spectral efficiency feedback is compared to the learned thresholds to either perform band switching, digital beam training, analog beam training, or data transmission. The second threshold is masked when the sub-6 GHz band is selected.
	\item \textbf{Greedy} policy: This algorithm chooses an action in each iteration following the genie-aided policy while being restricted to mmWave. This policy represents the performance that can be achieved with beam tracking and alignment alone, without the aid of a sub-6 GHz band.
\end{itemize}

%===========================================================%
\subsection{Numerical results and discussion} \label{sec:results_discussion}
%===========================================================%

\begin{figure}[t]
	\centering
	\includegraphics[width=0.6\columnwidth,draft=false]{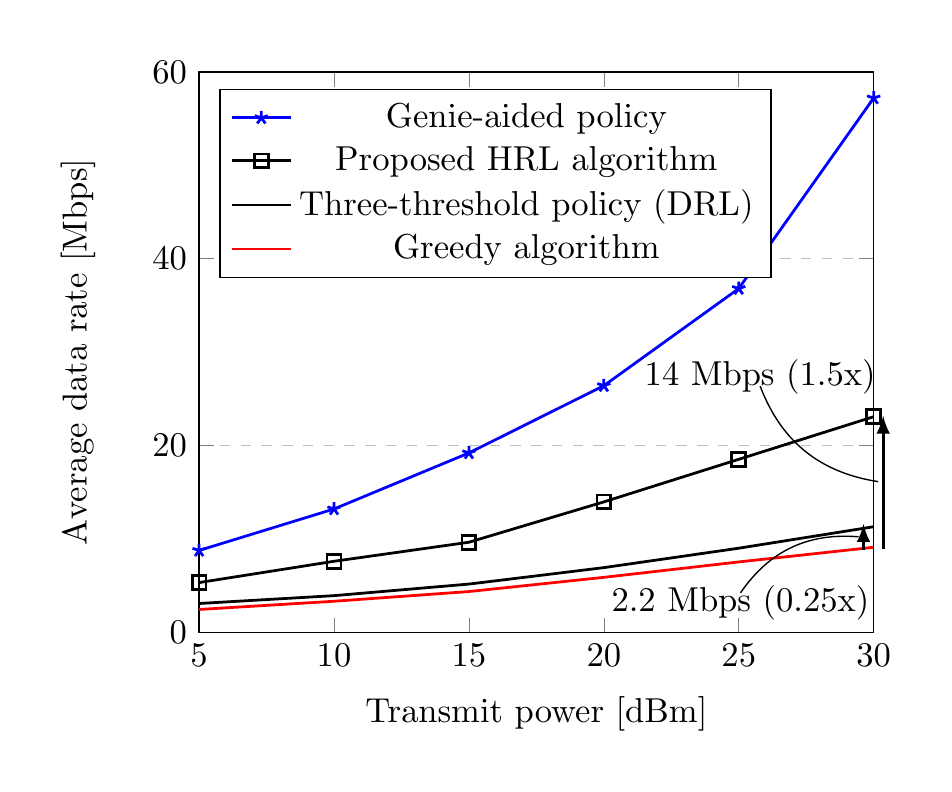}
	\caption{Illustration average data rate versus transmit power for (i) the genie-aided policy, (ii) the proposed HRL-based policy, (iii) DRL approach using threshold-based action, and (iv) the policy that only use the mmWave band. The distinctive beam management procedures between bands causes the DRL-based heuristic to incrementally improve over the greedy policy. Employing hierarchy between the band assignment and beam management leads to further improvement in the achieved data rate by resolving the nonstationary actions.}
	\label{fig:datarate-comparison}
\end{figure}

\figref{fig:datarate-comparison} shows the average data rate
versus transmit power, ranging over 5 dBm to 30 dBm. The proposed band assignment and beam management algorithm based on HRL outperforms the traditional DRL-based heuristic. At a high transmit power of 30 dBm, the HRL-based algorithm shows a 2.7-fold improvement over the greedy method in contrast to the DRL-based heuristic getting 0.25-fold gain over the greedy baseline. This suggests that the HRL-based method effectively learns the policy by decomposing the joint band assignment and beam management, unlike the DRL approach, which struggles with the nonstationary action between the sub-6 GHz and mmWave band.

% HRL-based approach achieves a data rate of 23.1 Mbps, while the DRL-based heuristic and greedy method achieve data rates of 11.3 Mbps and 9.1 Mbps, respectively. The

\begin{figure}[t]
	\centering
	\includegraphics[width=0.6\columnwidth,draft=false]{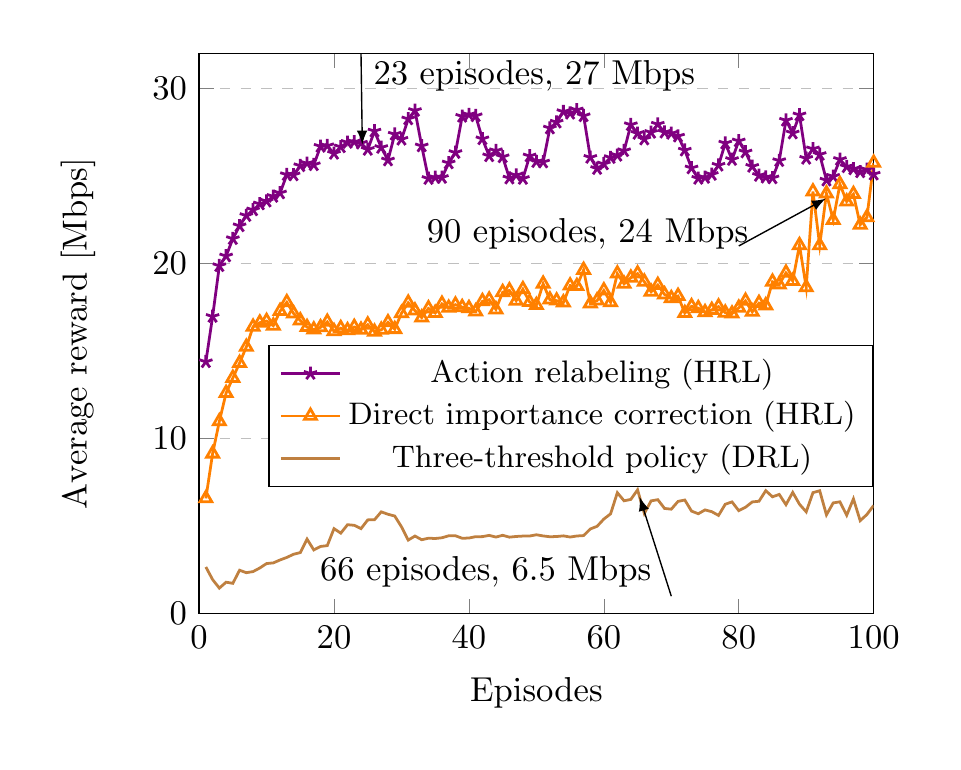}
	\caption{Reward convergence over learning episodes compared between the HRL algorithm with importance-based action relabeling, the HRL algorithm with direct importance correction, and the DRL algorithm with no hierarchy. Both HRL algorithms enjoy steep increase of average reward over the DRL approach. With different off-policy correction methods, action relabeling is beneficial in faster convergence whereas direct importance correlation is better in less deviation of reward.}
	\label{fig:sumrate-convergence}
\end{figure}

\figref{fig:sumrate-convergence} displays a comparison of the achieved data rate over 100 training episodes between the proposed HRL-based algorithm and the traditional RL algorithm as a baseline. Additionally, we implement direct importance correction as a baseline to examine its impact on the algorithms' performance. The results demonstrate that both HRL algorithms outperform the DRL approach, exhibiting a substantial increase in average reward. Among the different off-policy correction methods, action relabeling promotes faster convergence, while direct importance correction results in less deviation of reward. The DRL-based method takes around 60 episodes to converge at approximately 6.5 Mbps, whereas the HRL algorithms can achieve up to 27 Mbps. Notably, the importance-based action relabeling leads to the fastest convergence in approximately 20 episodes, while the direct importance correction method takes around 90 episodes to achieve more than 24 Mbps. We observed hours of runtime using a simulation environment with a GTX 1080 GPU to achieve the 27 Mbps of the HRL algorithm throughout 20 episodes. Still, base station deployments typically last for tens of years. This indicates that the investment of time in training is justified by the long-term performance benefits.

% Using a GPU of GTX 1080 Ti, a single iteration took around 3ms. Real world deployment takes around 10~20 hours to include 1000 channel sample average and 100 episode, each with length 200 iterations. The deployment will take tens of hours to train, though typical base stations will last for tens of years with sufficient maintanence and power supply.

\begin{figure}[t]
	\centering
	\includegraphics[width=0.6\columnwidth,draft=false]{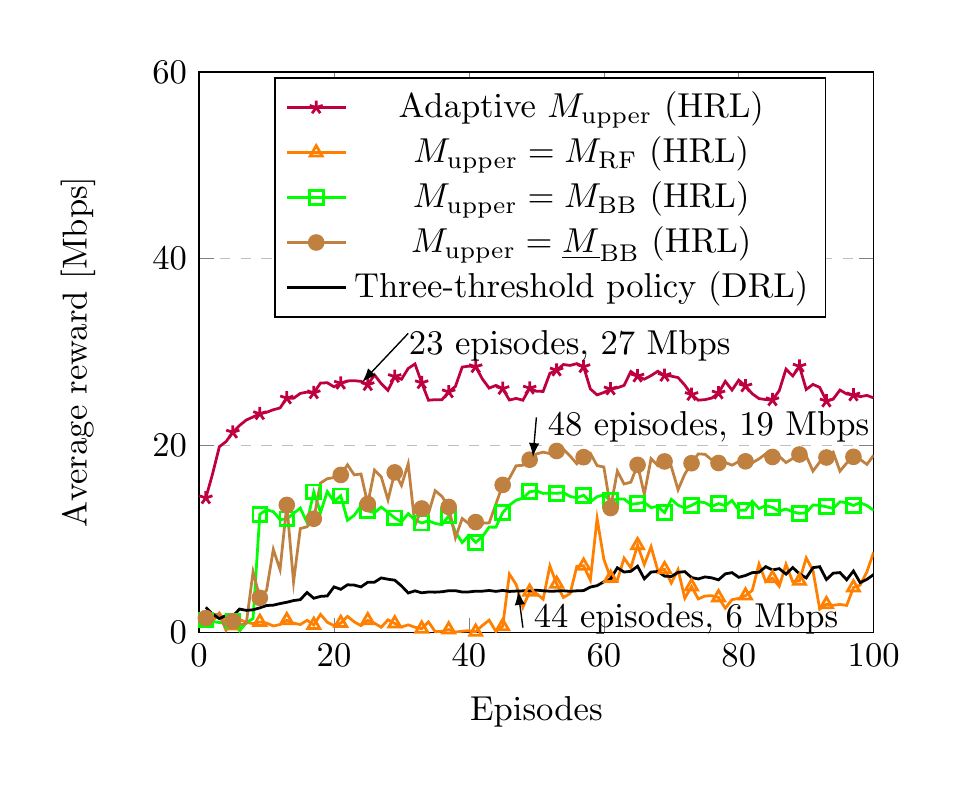}
	\caption{Convergence behavior over the period of the upper-level policy. The proposed method that uses round-skipping outperforms the baselines that have a fixed period of upper-policy trajectory sampling. The performance of the fixed period policies can deteriorate down to the vanilla DRL appraoch, losing the benefits of using the hierarchical structured learning.}
	\label{fig:upper-policy-period}
\end{figure}

\figref{fig:upper-policy-period} displays a comparison of the achieved data rate over 100 training episodes between HRL algorithms with different approaches to set the upper-level policy period. The proposed method based on round-skipping that adapts the upper-level period shows the best convergence behavior, converging to 27 Mbps around 20 episodes. While moderately short fixed upper-level period shows convergence around 20 episodes as well, the achieved reward drops down below 20 Mbps. When the fixed upper-level period is excessive, matching the analog beam training overhead, the performance is comparable to the vanilla DRL approach taking over 40 episode to converge to the reward below 10 Mbps. This highlights the importance of adaptively adjusting the upper-policy trajectory sampling period in achieving better performance in the HRL setting, where round-skipping is an effective way to adjust the upper-policy period.

\begin{figure}[t]
	\centering
	\includegraphics[width=0.6\columnwidth,draft=false]{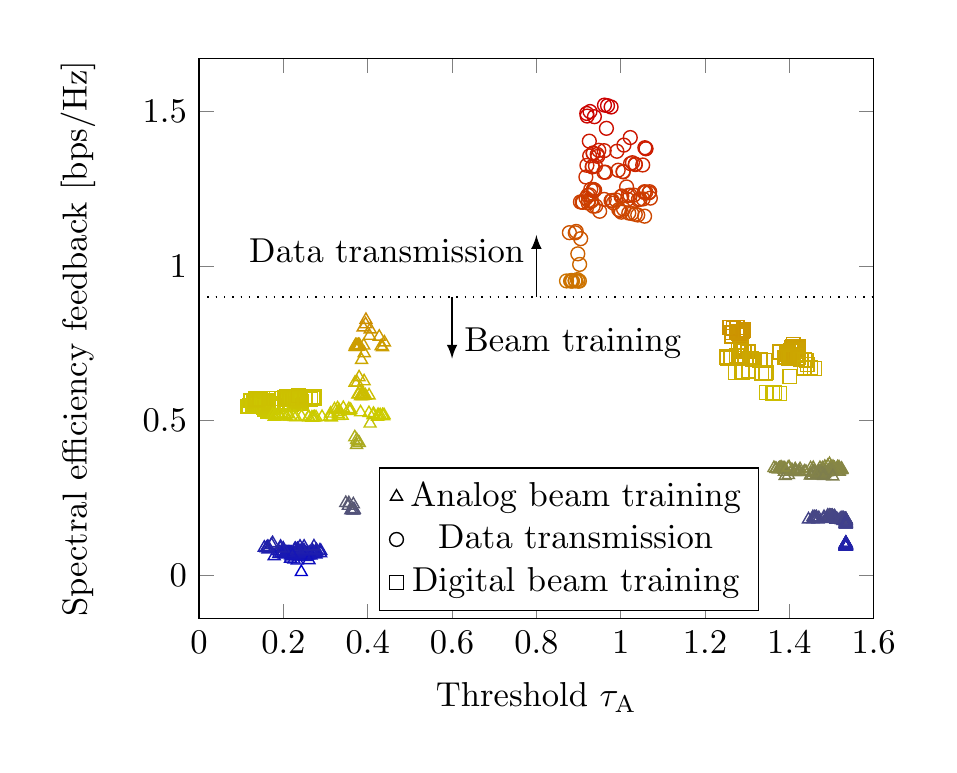}
	\caption{Beam training behavior per the learned threshold and spectral efficiency feedback at the mmWave band. Three clusters are observed: data transmission, analog beam training, and digital beam training. Clusters form based on spectral efficiency feedback and threshold level, with high spectral efficiency feedback for data transmission, moderate spectral efficiency feedback and high threshold for digital beam training, and low spectral efficiency feedback for analog beam training.}
	\label{fig:beam-training-pattern}
\end{figure}

\figref{fig:beam-training-pattern} shows the threshold and spectral efficiency feedback datapoints of the proposed HRL-based algorithm at the mmWave band during episodes 1-40 of the learning phase. Clusters are evident, with data transmission occurring when spectral efficiency feedback is over 0.9 bps/Hz and the threshold is between 0.8 and 1.2. Beam training occurs when the feedback deteriorates, with a high threshold indicating digital beam training and a low threshold indicating analog beam training. The cluster formation indicates that the threshold-based action in the HRL algorithm enables efficient beam management per spectral efficiency feedback.

\begin{figure}[t]
	\centering
	\includegraphics[width=0.6\columnwidth,draft=false]{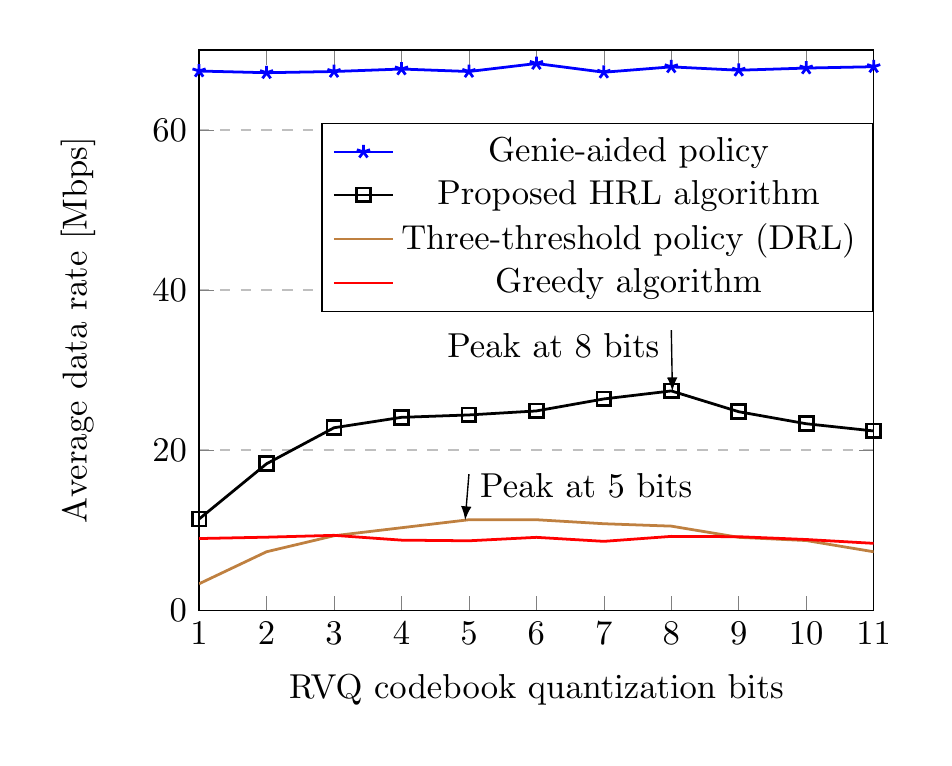}
	\caption{Achievable rate per the number of bits used in the digital effective channel quantization. Learning-based methods show a peaked curve in performance, with data rate increasing at low quantization due to more accurate channel estimates, but decreasing at excessive quantization due to significant overhead. The proposed HRL-based method outperforms the baseline DRL-based method, with the HRL-based method saturating at 8 bits and the DRL-based method at 5 bits.}
	\label{fig:RVQ-quantization}
\end{figure}

\figref{fig:RVQ-quantization} shows the average data rate per the number of quantization bits of the RVQ codebook used in the digital effective channel feedback. The range of the bits is selected from 1 through 11. Increasing the quantization codebook bits from 1 shows an increase in the average data rate since the digital effective channel feedback will become more accurate. The increase in the average data rate continues up to the quantization bits of 5 for the three-threshold policy and 8 for the proposed HRL-based method. We interpret that the DRL approach is bound to the number of SS blocks $\NSS=4$ whereas the HRL-based method can further benefit from the accurate digital effective channel to increase the achievable rate. Moreover, the propose HRL-based algorithm outperforms the greedy approach over the codebook quantization bits ranging from 1 through 11, where for the three-threshold policy the quantization codebook is preferred within 4 to 8 to outperform the greedy baseline.

\begin{figure}[t]
	\centering
	\includegraphics[width=0.6\columnwidth,draft=false]{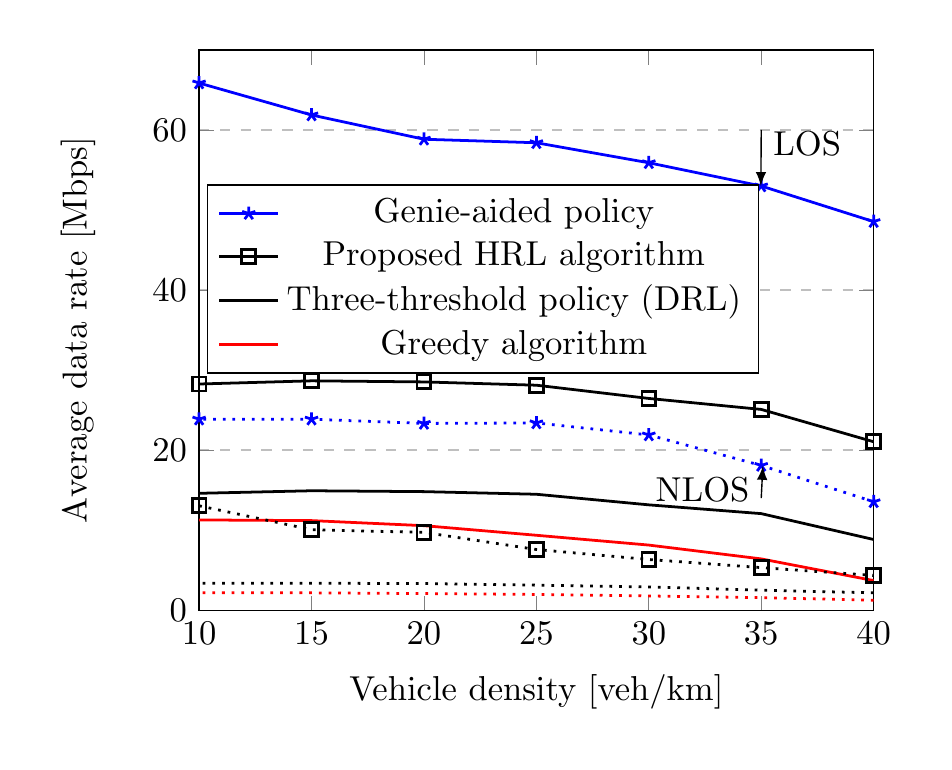}
	\caption{Achievable rate per different vehicle densities. The solid line shows the 3GPP-UMi LOS scenario and the dotted line shows the 3GPP-UMi NLOS scenario. Although increased vehicle density in urban scenarios can reduce the achievable data rate due to more frequent blockages, the HRL-based method can exploit the LOS channel to achieve milder performance loss compared to the NLOS scenario.}
	\label{fig:Vehicle-density}
\end{figure}

\figref{fig:Vehicle-density} shows the average data rate achievable per vehicle density, ranging from 10 to 40 vehicles per kilometer in the SUMO simulation, under different QuaDRiGa scenarios. The solid lines represent the performance of the policies under the 3GPP-UMi LOS scenario, while the dotted lines depict the performance of the policies under the 3GPP-UMi NLOS scenario. As the vehicle density increases, resulting in a higher likelihood of blockages, the achievable data rate decreases. However, our observation reveals that the exploitation of the LOS channel in the HRL-based method experiences a comparatively lesser performance loss in contrast to the NLOS scenario. The performance of the proposed HRL algorithm outperforms the vanilla DRL approach and the greedy algorithm under the LOS scenario over the increasing vehicle density, where we observe the comparable degradation of policy performance. The proposed method also outperforms the baselines under the NLOS scenario, but the performance degradation is more severe under the NLOS scenario over the increasing vehicle density. This observation may be due to the fact that the Type-1 codebook in the sub-6 GHz band is designed for LOS conditions to enjoy the short beam training overhead while maintaining high data rate. An interesting future direction in this regard would be to consider a Type-2 codebook with more sophisticated precoder computation that accounts for the multipath channel in the sub-6 GHz band.

%%%%%%%%%%%%%%%%%%%%%%%%%%%%%%%%%%%%%%%%%%%%%%%%%%%%%%%%%%%%%
\section{Conclusions and future work} \label{sec:conclusion}
%%%%%%%%%%%%%%%%%%%%%%%%%%%%%%%%%%%%%%%%%%%%%%%%%%%%%%%%%%%%%

Exploiting multiple band characteristics will be a major approach addressing challenges in wireless networks, including mobility and blockage, while avoiding their associated drawbacks. We formulated the joint band assignment and beam management problem in wireless networks operating on FR1 and FR2. We devised an MDP that introduces hierarchy between the band assignment and beam management to avoid nonstationary action space. The numerical evaluation based on QuaDRiGa-generated channel showed that the proposed HRL-based method improves over traditional DRL approaches. This suggests that the introduction of hierarchy is an effective approach addressing the complex problem of joint band assignment and beam management. For future work, the extension to multi-user scenario is an interesting direction that may require queuing theory to resolve conflicts between users with the same preferred band.

\bibliography{Dohyun_IEEE_reference}	
\bibliographystyle{IEEEtran}

\end{document}

%% file: Dohyun_input.tex
\usepackage{datetime}
\usepackage{color,soul,cprotect}
\usepackage{mathtools}
\usepackage{booktabs}
\usepackage{multirow} 
\usepackage{cite}
\usepackage{amssymb} 
\usepackage{amsfonts}
\usepackage{amsthm}
\usepackage{verbatim,tikz,pgfplots}

\usepackage{dblfloatfix}

\usetikzlibrary{automata,positioning}
\usetikzlibrary{decorations.pathreplacing,positioning, arrows.meta}

\usepackage{ctable,color,colortbl}
\usepackage{graphicx,epstopdf}
\usepackage{setspace}
\usepackage{algorithm}
\usepackage{algorithmic,eqparbox}

\usepackage{enumerate}
\usepackage[margin=1 in]{geometry}
\usepackage{hyperref}
\usepackage{bbm,bm}
\usepackage{array}
\newcolumntype{?}{!{\vrule width 1pt}}

\usepackage[caption=false,font=normalsize,labelfont=sf,textfont=sf]{subfig}
\usepackage{color}

\makeatletter
\newcommand{\biggg}{\bBigg@{2}}
\newcommand{\Biggg}{\bBigg@{3}}
\newcommand{\Bigggg}{\bBigg@{4}}
\makeatother
\DeclareMathOperator*{\argmax}{argmax}
\DeclareMathOperator*{\argmin}{argmin}

\newcommand{\figref}[1]{{Fig.}~\ref{#1}}

\newcommand{\secref}[1]{{Section}~\ref{#1}}
\newcommand{\algref}[1]{{Algorithm}~\ref{#1}}

\def\bb0{{\mathbb{0}}}

\def\ba{{\mathbf{a}}}
\def\bb{{\mathbf{b}}}

\def\b0{{\mathbf{0}}}

\def\bA{{\mathbf{A}}}

\def\bI{{\mathbf{I}}}

\def\bbE{{\mathbb{E}}}

\def\cA{\mathcal{A}}

\def\cC{\mathcal{C}}
\def\cD{\mathcal{D}}

\def\cF{\mathcal{F}}

\def\cH{\mathcal{H}}

\def\cN{\mathcal{N}}

\def\cT{\mathcal{T}}

\def\cW{\mathcal{W}}

\def\sf0{{\mathsf{0}}}

\usepgfplotslibrary{external}

\newcommand{\mytilde}{\raise.17ex\hbox{$\scriptstyle\mathtt{‌​\sim}$}}

\graphicspath{{figure/}{figures/}}

\newcommand{\WRF}{\bm{\mathsf{W}}_{\text{RF}}}

\newcommand{\horizonlen}{M}
\newcommand{\horizonindex}{m}
\newcommand{\subcarriernum}{K}
\newcommand{\subcarrierindex}{k}

\newcommand{\modeindex}{n_{\text{mode}}}
\newcommand{\NS}{N_{\text{S}}}

\newcommand{\channelmatrix}{\bm{\mathsf{H}}}

\newcommand{\state}{\cT}
\newcommand{\action}{\cA}

\newcommand{\noiseSTD}{\sigma_{\text{n}}}

\newcommand{\thetaA}{\boldsymbol{\theta}_{\text{A,ON}}}
\newcommand{\thetaC}{\boldsymbol{\theta}_{\text{C,ON}}}
\newcommand{\targthetaA}{\boldsymbol{\theta}_{\text{A,TAR}}}
\newcommand{\targthetaC}{\boldsymbol{\theta}_{\text{C,TAR}}}

\newcommand{\NSS}{N_{\text{SS}}}
\newcommand{\SSperiodicity}{\horizonlen_{\text{SS}}}

\newcommand{\DTlen}{\horizonlen_{\text{DT}}}

\newcommand{\batchsize}{\xi}

\newcommand{\FRF}{\bm{\mathsf{F}}_{\text{RF}}}

\newcommand{\cmWavesubcarriernum}{\underline{K}}
\newcommand{\mmWavesubcarriernum}{K}
\newcommand{\cmWavebandwidth}{\underline{B}}
\newcommand{\mmWavebandwidth}{B}

\newcommand{\NTXRF}{N_{\text{TX},\text{RF}}}

\newcommand{\NBSRF}{N_{\text{BS},\text{RF}}}
\newcommand{\Nbs}{N_{\text{BS}}}
	
\newcommand{\userNr}{N_{\text{UE}}}
\newcommand{\userNRF}{N_{\text{UE},\text{RF}}}
\newcommand{\userNS}{N_{\text{S}}}
\newcommand{\usersymbol}{\mathsf{s}}
\newcommand{\userFBB}{\bm{\mathsf{F}}_{\text{BB}}}
\newcommand{\userWRF}{\bm{\mathsf{W}}_{\text{RF}}}
\newcommand{\userWBB}{\bm{\mathsf{W}}_{\text{BB}}}
\newcommand{\userchannel}{\bm{\mathsf{H}}}
\newcommand{\userEffectiveChannel}{\bar{\bm{\mathsf{H}}}}
\newcommand{\userQuantizedChannel}{\hat{\bm{\mathsf{H}}}}
\newcommand{\userlargescalefading}{G[\horizonindex]}
\newcommand{\usernoise}{\bm{\mathsf{n}}}
\newcommand{\usercodebook}{\cW}
\newcommand{\userTransmitpower}{P[\subcarrierindex,\horizonindex]}

\newcommand{\userRate}{R}
\newcommand{\userSE}{S}

\newcommand{\bandassign}{b}

\newcommand{\cmWavechannelmatrix}{\underline{\bm{\mathsf{H}}}}

\newcommand{\cmWaveNt}{\underline{N}_{\text{BS}}}

\newcommand{\cmWaveNS}{\underline{N}_{\text{S}}}
\newcommand{\cmWaveuserNr}{\underline{N}_{\text{UE}}}

\newcommand{\cmWaveuserNS}{\underline{N}_{\text{S}}}
\newcommand{\cmWaveuserlargescalefading}{\underline{G}[\horizonindex]}
\newcommand{\cmWaveuserFBB}{\underline{\bm{\mathsf{F}}}_{\text{BB}}}

\newcommand{\cmWaveuserWBB}{\underline{\bm{\mathsf{W}}}_{\text{BB}}}
\newcommand{\cmWaveuserchannel}{\underline{\bm{\mathsf{H}}}}

\newcommand{\cmWaveuserSE}{\underline{S}}

\newcommand{\cmWaveusernoise}{\bm{\underline{\mathsf{n}}}}
\newcommand{\cmWavenoiseSTD}{\underline{\sigma}_{\text{n}}}
\newcommand{\cmWaveuserTransmitpower}{\underline{P}[\subcarrierindex,\horizonindex]}

\newcommand{\Policy}{\mu}
\newcommand{\upperPolicy}{\Policy^{\text{upper}}}
\newcommand{\lowerPolicy}{\Policy^{\text{lower}}}
\newcommand{\lowerbaselinePolicy}{\Policy^{\text{lower}}_\text{base}}
\newcommand{\upperAction}{g}
\newcommand{\upperQ}{Q^{\text{upper}}}
\newcommand{\lowerQ}{Q^{\text{lower}}}
\newcommand{\lowerExperienceReplay}{\cD_{\text{lower}}}
\newcommand{\upperExperienceReplay}{\cD_{\text{upper}}}

\newcommand{\aggUpperState}{\state^{\text{agg}}}

\newcommand{\aggGoal}{\upperAction^{\text{agg}}}
\newcommand{\aggUpperReward}{\envReward^{\text{agg}}}

\newcommand{\precoderInfo}{\bm{\mathsf{P}}}

\newcommand{\sfg}{\bm{\mathsf{g}}}
\newcommand{\sfv}{\bm{\mathsf{v}}}

\newcommand{\TXcodebooksize}{\nu_{\text{BS}}}
\newcommand{\RXcodebooksize}{\nu_{\text{UE}}}

\newcommand{\TXindexRF}{n_{\text{BS},\text{RF}}}
\newcommand{\RXindexRF}{n_{\text{UE},\text{RF}}}

\newcommand{\BTlen}{\horizonlen_{\text{BT}}}

\newcommand{\AnalogBTOverhead}{M_{\text{RF}}}
\newcommand{\DigitalBTOverhead}{M_{\text{BB}}}
\newcommand{\cmWaveBTOverhead}{\underline{M}_{\text{BB}}}

\newcommand{\userRFSEFeedback}{S_{\text{UE}}}
\newcommand{\cmWaveuserRFSEFeedback}{\underline{S}_{\text{UE}}}

\newcommand{\intrinsicReward}{r_{\text{I}}}
\newcommand{\envReward}{r_{\text{E}}}
\newcommand{\upperPeriod}{\horizonlen_{\text{upper}}}
\newcommand{\importanceWeight}{w[\horizonindex]}
\newcommand{\relabedGoal}{\bar{\upperAction}}
\newcommand{\actAvailProb}{q(\action,\horizonindex)}

\newcommand{\DRLanalogAction}{\tau_{\text{RF}}}
\newcommand{\DRLdigitalAction}{\tau_{\text{BB}}}
\newcommand{\HRLanalogAction}{\tau_{\text{A}}}
\newcommand{\HRLdigitalAction}{\tau_{\text{D}}}

\newcommand{\analogPilotRatio}{\beta_{\text{RF}}}
\newcommand{\digitalPilotRatio}{\beta_{\text{BB}}}
\newcommand{\cmWavePilotRatio}{\underline{\beta}}
\newcommand{\analogOFDMFrames}{\zeta_{\text{RF}}}
\newcommand{\digitalOFDMFrames}{\zeta_{\text{BB}}}
\newcommand{\cmWaveOFDMFrames}{\underline{\zeta}}

\newcommand{\RVQbits}{\kappa_{\text{RVQ}}}
\newcommand{\channelbits}{\kappa_{\text{channel}}}
\newcommand{\PMIcodebooksize}{\nu_{\text{PMI}}}
\newcommand{\cmWavechannelbits}{\underline{\kappa}_{\text{channel}}}